\documentclass[table, screen, sigconf]{acmart}

\AtBeginDocument{%
  }

\setcopyright{rightsretained}
\copyrightyear{2027}
\acmYear{2027}
\acmDOI{XXXXXXX.XXXXXXX}

\acmConference[Conference acronym 'XX]{Make sure to enter the correct
  conference title from your rights confirmation emai}{June 03--05,
  2027}{Woodstock, NY}
\acmISBN{978-1-4503-XXXX-X/18/06}

\usepackage[flushleft]{threeparttable}
\usepackage{multirow}
\usepackage{pdflscape}
\usepackage{subcaption}
\usepackage{caption}
\usepackage{array}
\usepackage{colortbl} 
\usepackage{comment}
\usepackage{balance}
\usepackage{longtable}
\usepackage{booktabs}
\usepackage{multirow}
\usepackage{ragged2e}
\usepackage{float}
\usepackage{subcaption}

\usepackage{arydshln}
\usepackage[table]{xcolor}
\definecolor{lightgray}{gray}{0.90}
\definecolor{headergray}{gray}{0.80}

\begin{document}

\title[A System of Care, Not Control]{A System of Care, Not Control: Co-Designing Online Safety and Wellbeing Solutions with Guardians ad Litem for Youth in Child Welfare}

%Bridging Advocacy and Design: Co-Creating Online Safety Tools with Guardians ad Litem OR Supporting Advocacy Through Design: Addressing GALs’ Challenges in Promoting Youth Online Safety


\author{Johanna Olesk}
\orcid{0009-0003-1707-0750}
\affiliation{%
\department{Computer Science and Engineering}
  \institution{University of Notre Dame}
  \city{Notre Dame}
  \state{Indiana}
  \country{USA}}
\email{jolesk@nd.edu}

\author{Ozioma C. Oguine}
\orcid{0000-0003-2434-1400}
\affiliation{%
\department{Computer Science and Engineering}
  \institution{University of Notre Dame}
  \city{Notre Dame}
  \state{Indiana}
  \country{USA}}
\email{ooguine@nd.edu}

\author{Mariana Fernandez Espinosa}
\orcid{0009-0004-1116-2002}
\affiliation{%
\department{Computer Science and Engineering}
  \institution{University of Notre Dame}
  \city{Notre Dame}
  \state{Indiana}
  \country{USA}}
\email{mferna23@nd.edu}

\author{Alexis B. Peirce Caudell}
\orcid{0009-0001-2115-1989}
\affiliation{%
\department{Luddy School of Informatics, Computing, and Engineering}
  \institution{Indiana University}
  \city{Bloomington}
  \state{Indiana}
  \country{USA}}
\email{abpeirce@iu.edu}

\author{Karla Badillo-Urquiola}
\orcid{0000-0002-1165-3619}
\affiliation{%
\department{Computer Science and Engineering}
  \institution{University of Notre Dame}
  \city{Notre Dame}
  \state{Indiana}
  \country{USA}}
\email{kbadillou@nd.edu}

\renewcommand{\shortauthors}{Olesk et al.}

\begin{abstract}
    % \textcolor{red}{1 sentence background and context (intro and problem)}\\
    % \textcolor{red}{1 sentence gap/problem statement}\\
    % \textcolor{red}{1 sentence research focus/methods}\\
    % \textcolor{red}{2-3 sentences key findings/arguments}\\
    % \textcolor{red}{1-2 sentences contributions/implications}

    Current online safety technologies overly rely on parental mediation and often fail to address the unique challenges faced by youth in the Child Welfare System (CWS). These youth depend on a complex ecosystem of support, including families, caseworkers, and advocates, to safeguard their wellbeing. Within this network, Guardians ad Litem (GALs) play a unique role as court-appointed advocates tasked with ensuring the best interests of youth. Yet little is known about how GALs perceive and support youths’ online safety. To address this gap, we conducted a two-part workshop with 10 GALs to explore their perspectives on online safety and collaboratively envision technology-based solutions tailored to the needs of youth in the CWS. Our findings revealed that GALs struggle to support youth with online safety challenges due to limited digital literacy, inconsistency of institutional support, lack of collaboration among stakeholders, and complexity of family dynamics. While GALs recognized the need for some oversight of youth online activities, they emphasized designing systems that support online safety beyond control or restriction by fostering stability, trust, and meaningful interactions, both online and offline. GALs emphasized the importance of developing tools that enable ongoing communication, therapeutic support, and coordination across stakeholders. Proposed design concepts focused on strengthening youth agency and cross-stakeholder collaboration through virtual avatars and mobile apps. This work provides actionable design concepts for strengthening relationships and communication across care network. It also redefines traditional approaches to online safety, advocating for a holistic, multi-stakeholder online safety paradigm for youth in the CWS.

    %Conventional online safety technologies overly rely on parental controls and oversight, yet these approaches often fall short for youth in the Child Welfare System (CWS), who depend on complex ecosystems of support—including caseworkers, therapists, caregivers, and advocates—for their wellbeing. Guardians ad Litem (GALs), as court-appointed advocates, play a critical role in representing the best interests of youth and helping coordinate care across fragmented systems. Yet their perspectives on online safety remain largely overlooked, even though they are uniquely positioned to observe how systemic gaps, fractured relationships, and inconsistent digital mediation shape youth’s online vulnerabilities. We conducted a co-design workshop with ten GALs to understand their perspectives and explore more effective, contextually grounded online safety solutions. Our findings reveal that GALs face significant challenges in advocating for youth digital safety, including systemic barriers, limited digital literacy, and fragmented communication. Rather than favoring surveillance or restriction, GALs envisioned trust-based, therapeutic, and collaborative solutions, such as a digital avatar to support emotional regulation and a collaborative communication platform to bridge stakeholders. This work contributes actionable design directions that reframe online safety as a collective, trauma-informed responsibility grounded in relational care.
    
\end{abstract}






\maketitle

\section{Introduction}

Guardians ad Litem (GALs) play a critical role in the U.S. child welfare system (CWS), serving as court-appointed advocates tasked with representing the best interests of children who have experienced abuse, neglect, or other forms of harm \cite{The_GAL_2019, FAUHow2025Gal}. According to a 2023 ACARS report, the U.S. child welfare system served an estimated 546,159 children who were victims of abuse and neglect, with approximately 343,077 of these children placed in foster care \cite{AFcars}. In these contexts, GALs act as mediators and advisors, building relationships with children, investigating their circumstances, and making recommendations to the court about decisions that would best support the child’s overall well-being.

Traditionally, GALs focus on advocating for the child's physical safety, mental and emotional health, and healthy family dynamics \cite{GAL_Role_2012, FAUHow2025Gal}. However, as digital technologies become increasingly pervasive in youths' lives, questions arise about how GALs can support children's online well-being. Youth in the CWS are among the most vulnerable to online risks, given their histories of trauma, instability, and fractured support systems \cite{badillo2019risk, caddle2022challenge, oguine2025Chins}. Research has shown that given the vital role online technologies provide youth in the CWS to connect with peers and seek support or education, they face higher risks of encountering online harms such as grooming, cyberbullying, exposure to harmful content, and exploitation compared to their peers \cite{badillo2019risk, oguine2025Chins}.

In response, many child welfare practices have leaned toward restrictive approaches, such as limiting or even prohibiting youth's access to technology in an effort to shield them from harm \cite{badillo2024caseworker}. However, these restrictive measures can also isolate them further and deny them access to social support, education, and empowerment opportunities that digital spaces offer their peers \cite{badillo2019risk}. This tension, as pointed out by \citet{badillo2019risk}, raises an important question: \textit{How can GALs effectively advocate for the online safety of youth in the CWS while respecting their need for connection, support, and autonomy in digital spaces?} To address this gap, we sought to understand how GALs perceive the online safety challenges youth in the CWS face, how they currently advocate for online safety, and what solutions they envision to better support youth in navigating online risks. Specifically, we propose the following research questions:

\begin{itemize}
    \item \textbf{RQ1:} According to GALs, what online safety challenges do families seek their support for?
    \item \textbf{RQ2:} What challenges do GALs face when advocating for the online safety of youth?
    \item \textbf{RQ3:} What technology-based solutions do GALs envision for improving the online safety of youth in the child welfare system?
\end{itemize}

To answer these questions, we conducted a two-part workshop with (N=10) GALs. Part I of the workshop was dedicated to content delivery and reflective discussion, while in Part II was a co-design activity. Using thematic analysis of the workshop discussions and co-design artifacts, we explored both the challenges families and GALs face regarding youth online safety, and the technology-based solutions GALs envisioned to address these issues. We found three key insights. First, GALs observed that youth in the child welfare system face heightened emotional vulnerabilities and social isolation, which, combined with inconsistent family rules and support, leave them particularly exposed to online risks. Second, GALs themselves encounter systemic barriers, fragmented communication among stakeholders, and gaps in digital literacy that hinder their ability to advocate effectively for youth’s online safety. Finally, through co-design activities, GALs envisioned technology-based solutions that move beyond restriction and surveillance, emphasizing trust-building, youth agency, coordinated stakeholder communication, and integration of offline and therapeutic support into digital tools.

Building on these insights, we reconceptualize online safety as a value of relational care, recognizing its inseparability from emotional and relational wellbeing. We further argue that online safety must be understood as a multi-stakeholder responsibility rather than an individual obligation, advocating for a collective care approach to supporting youth in the CWS as they navigate digital environments. Finally, we outline three key implications for HCI and GROUP communities: designing for multi-stakeholder collaboration, trust-based communication, and therapeutic relational support with online-to-offline transitions to better foster youth digital safety and wellbeing. Our research makes three main contributions to the HCI and GROUP communities:

\begin{itemize}
    \item We provide the first empirical account of GALs' perspectives on youth online safety in the CWS, illuminating the unique online challenges they observe among families and the systemic and relational barriers GALs encounter in advocating for youth digital wellbeing
    \item We demonstrate the value of a co-design for engaging underexplored yet critical stakeholders, such as GALs, in envisioning technology-based interventions.
    \item We articulate design considerations for online safety interventions that go beyond surveillance and restriction, emphasizing trust-building, youth agency, coordinated stakeholder communication, and integration of therapeutic and offline support.
\end{itemize}

% \textcolor{red}{}
% - start with how Guardians ad Litem in the US are key stakeholders in CWS, advocating for the best interests of the child in the system\\
% - They serve as mediators and advocates, making suggestions to the court about what decisions are in the best interests of the child\\
% - conventionally, GALs must advocate for the physical wellbeing, mental wellbeing and healthy family dynamics for the child, but what about the online wellbeing?\\
% - with the increasingly pervasive online use by youth, GALs need to also be ready to advocate for the online safety of youth\\
% - Youth in the CWS are highly vulnerable to online risks (cite research)\\
% - therefore, most approaches have taken restrictive approaches (show stats of their access to online/tech?)\\
% - so how do they advocate for the online safety of in these conditions?\\
% - We aimed to understand how GALs perceive the online safety of youth in the CWS and how they advocate for it.\\
% - our research questions\\

% - To answer these questions, we conducted a one-day workshop with 10 GALs in Indiana\\
% - We used thematic analysis to understand the challenges families face regarding youth online safety and the challenges GALs face when advocating for youth online safety. Also what solutions they envision\\
% - Our findings show that...\\
% - We propose ...\\
% - Our contributions are three-fold\\

\section{Background}

To situate our study, we first reviewed HCI research on technology in the child welfare system, highlighting how it has addressed the needs of youth and stakeholders. We then examined work on the online safety of youth in the CWS and the importance of participatory and co-design approaches for developing effective interventions. Finally, we described the unique role of GALs as advocates and their potential to inform youth online safety solutions.

\subsection{Investigating the Child Welfare System in HCI}
% --> This includes Devansh, Kenn, Trove platform, etc
% --> Memories, AI integration, etc.

Researchers in the SIGCHI community have previously examined the role of technology within the U.S. child welfare system (CWS), seeking to understand and design systems that support the well-being of children \cite{badillo2019risk, oguine2025Chins}, improve stakeholder practices \cite{saxena2020pd_study, badillo2024caseworker}, and address the challenges faced by both youth and practitioners \cite{badillo2024caseworker, fitch2012youth, gustavsson2015positive}. Prior GROUP and HCI scholarship on the child welfare system (CWS) has approached the topic through both technical and sociotechnical lenses, contributing insights into child welfare policies \cite{saxena2020group, fitch2012youth, gustavsson2015positive}, stakeholder practices \cite{badillo2024caseworker, badillo2019risk}, youth well-being \cite{badillo2019risk, oguine2023you}, and the implications of algorithmic decision-making for resource allocation \cite{saxena2023analysis, brown2019Algo, saxena2023public_sector}. For example, \citet{saxena2020pd_study} conducted participatory design to improve public-sector algorithms, revealing the difficulties of incorporating stakeholder input into algorithmic decision-making processes. In a follow-up study \cite{saxena2024street_level}, they uncovered how algorithmic systems in child welfare introduce harms and uncertainties stemming from organizational structures, street-level decision-making, and gaps in accountability. While these studies emphasize the importance of centering stakeholders in the design and deployment of technologies, they also suggest that the systemic constraints of the CWS can undermine even well-intentioned interventions if they do not account for youth realities.

Complementing these systemic and stakeholder-focused investigations, other studies have begun to foreground the experiences and agency of youth themselves. For example, \citet{oguine2025Chins} analyzed self-reported data from youth on a mental health support platform, revealing the complex ways they navigate digital spaces to cope with adverse childhood experiences, often encountering risks alongside support. Similarly, \citet{badillo2019risk, badillo2024caseworker} documented how foster youth’s online behaviors reflect their desire for autonomy and connection, even as caregivers and caseworkers struggle to mediate these activities effectively. These findings were corroborated by an interview study conducted by \citet{caddle2022challenge} with Social Service Providers (SSPs), which revealed that organizational policies, limited resources, and competing priorities often hinder stakeholders’ ability to support the online safety of underprivileged youth effectively. Together, these studies highlight both the vulnerabilities and the resilience of youth in the CWS, underscoring the need to incorporate their voices, along with those of stakeholders, into the design of technology-mediated interventions. Building on prior research, our study engaged GALs in a workshop to surface their insights and collaboratively envision technology-based solutions that promote youth online safety, contributing a novel stakeholder perspective to the growing literature on child welfare and HCI.

\subsection{Understanding the Online Safety of Youth in the Child Welfare System}
% Karla's work \& more...

Youth online safety has been extensively studied at CHI and the broader HCI community (e.g., \cite{wisniewski2016dear, livingstone2008taking, park2024personally, akter2025calculating, alsoubai2022mosafely, razi2023sliding, oguine2026define_safety, McNally2018codesign}), focusing on understanding the opportunities and risks youth face online, and developing safeguarding methods to protect them. While traditional approaches to online safety have largely focused on restrictive parental control and monitoring \cite{agha2021just, iftikhar2021designing, schiano2017parental}, recent scholarship has shifted towards empowerment- and resilience-based methods to provide youth with more autonomy and control over their online safety \cite{agha2023strike, hartikainen2016should, wisniewski2017parental}. 
%Researchers recognize the need for collaborative and transparent online risk management \cite{oecd2025childrendigital}.
For example, \citet{freed2025protect} developed the PROTECT framework, which explores how social, contextual, and dynamic factors shape help-seeking behaviors in the face of digital risks, highlighting the importance of collective, youth-centered approaches to fostering digital resilience. Similarly, \citet{agha2024tricky} engaged teens in co-design sessions to create and evaluate nudges for youth online risk prevention, emphasizing realistic and nuanced risk scenarios with nudges that empower youth. Yet, much of this literature remains largely based on research conducted with typical youth from affluent families and in Global North contexts \cite{oguine2025online, katieResiliance2024}. Scholars have noted that these limitations can lead to solutions that take on a "one-size-fits-all" approach to online safety, failing to account for the complexities and realities
of youth \cite{katieResiliance2024}. Instead, scholars advocate for moving towards more nuanced and targeted approaches for vulnerable populations to account for sociocultural, familial, and institutional contexts shaping youth’s digital experiences \cite{pinter2017adolescent, oguine2025online}. This approach is particularly important for understanding the experiences of youth in the child welfare system (CWS), where assumptions underlying typical parent–child mediation models may not always apply \cite{caddle2023duty, badillo2017understanding}.

Youth in the CWS, such as foster care, often experience emotional, physical, and social challenges, which exacerbate their vulnerability to online harms \cite{oguine2025Chins, badillo2019risk, badillo2024caseworker, fitch2012youth}. Researchers have investigate the online safety of youth in the CWS through various approaches, including stakeholder-centered studies capturing the perspectives of caregivers and professionals \cite{badillo2024caseworker, badillo2019risk}, self-reported accounts from youth \cite{oguine2025Chins}, and policy-oriented analyses \cite{fitch2012youth}. For instance, \citet{fitch2012youth} applied Critical Systems Heuristics to inform privacy policies for teens in foster care, recommending collaborative policy development with youth and other stakeholders that emphasizes empowerment, privacy, and education over restriction. Beyond policy, \citet{gustavsson2015positive} highlighted the potential of digital technologies to support positive youth development in foster care, advocating for digital literacy and supportive youth–adult partnerships to empower youth and balance safety with opportunities for connection and growth. In their interview studies with foster parents and caseworkers, \citet{badillo2019risk, badillo2024caseworker} examined how foster parents and caseworkers work together to mediate youth technology use, emphasizing the need to design tools that reflect the unique online safety needs of foster families. The authors advocate for collaborative, socio-technical, ecosystem-based support for caseworkers, foster parents, and youth to enable early intervention and safeguard offline well-being. More recently, \citet{oguine2025Chins} found that youth identifying as in need of services frequently encounter adverse online experiences when seeking support online, recommending tailored, trauma-informed support systems and technical solutions that foster empowerment, privacy education, and participatory design.

Prior research has highlighted the value of centering stakeholder and youth perspectives in understanding and addressing online safety in the CWS. We build on this work by providing the critical perspective of GALs. GALs occupy a unique position in the CWS: they have visibility into the broader systemic and interpersonal challenges youth face and hold the authority to make independent recommendations directly to the court. Building on these insights, our study engages GALs in a co-design activity to surface their experiences and collaboratively envision technology-based solutions to promote the online safety of youth in the CWS.

\subsection{The Role of Guardians ad Litem and Child Welfare Advocates}
% --> Description of what they do in the system and why they're essential for understanding online safety in cws.
%found this, might be good overview of GAL duties etc: https://avnetlaw.com/2018/06/20/guardian-ad-litem-indiana/

Guardians ad Litem (GALs) and Court-Appointed Special Advocates (CASAs) are court-appointed volunteers, often educators, attorneys, or retirees, who advocate for the best interests of children in abuse or neglect cases \cite{GAL_Role_2012, The_GAL_2019, Badillo2018stakeholder}. They conduct independent investigations, maintain ongoing relationships with youth, and advise the courts to help secure safe and permanent placements for children \cite{GAL_Role_2012}. Current research from Florida highlights how GALs build trusting and supportive relationships with foster youth, yet often face gaps in training, understanding, and resources when addressing online risks faced by these youth \cite{thompson2025forming}. Given their deep involvement with children and their systemic visibility within the child welfare process, GALs offer a critical perspective for designing online safety interventions that are both youth-informed and feasible within the bureaucratic realities of the CWS.

While co-design approaches within HCI have long demonstrated that involving key stakeholders in participatory processes leads to more contextually grounded and effective interventions \cite{kumar2025cultivating, zytko2022participatory, McNally2018codesign, badillo2019stranger}, youth in the CWS remain largely excluded from such processes due to ethical and logistical barriers \cite{badillo2017understanding}. Consequently, adults such as GALs often act as intermediaries who interpret, mediate, and support youths’ digital lives \cite{badillo2024caseworker, cheng2022Workers}. Yet, this group has been largely overlooked in prior design research, despite their close understanding of both the systemic constraints and the emotional realities youth face. Engaging GALs in co-design not only surfaces their unique expertise at the intersection of youth advocacy and legal decision-making but also ensures that proposed online safety solutions are both actionable within the child welfare system and aligned with the relational and emotional realities of youth in care. Our work addresses this gap by engaging GALs in a participatory workshop, highlighting their perspectives to inform equitable, trauma-informed, and system-aware interventions that promote youth online safety within the child welfare practice.
\section{Methods}

This section describes the study design (including workshop facilitation and activities), participant recruitment and demographics, and qualitative data analysis.

\subsection{Study Overview}

The goal of this study was threefold: (1) to inform GALs about the latest youth online safety research and safeguarding strategies; (2) to understand the unique challenges that youth in the child welfare system (CWS) face online, as well as the challenges GALs encounter when advocating for their online safety; and (3) to collaboratively design technology-based solutions to promote the online safety of youth in the CWS. Guided by the principles of participatory design \cite{kensing1998participatory}, we directly engaged active GALs to center their perspectives on the online safety of youth in the child welfare system and to surface the challenges they encounter while advocating for these youth. We chose a workshop approach with discussion and co-design activities because GALs possess unique, practice-based expertise that is crucial for developing feasible and contextually appropriate online safety solutions. 

\begin{figure*}[!t]
  \centering
  \begin{subfigure}{0.32\textwidth}
    \centering
    \includegraphics[width=\linewidth]{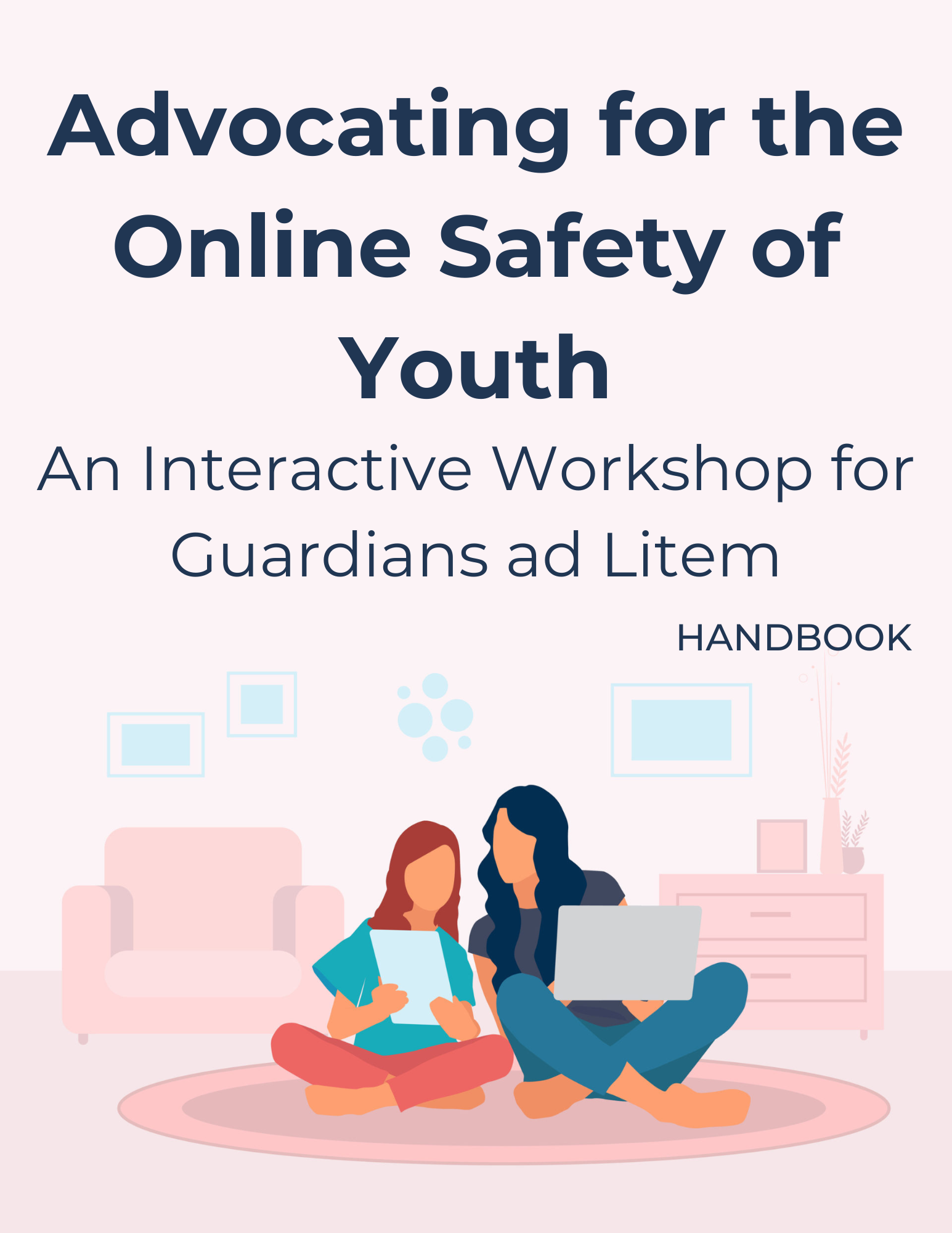}
    \label{fig:handbookfirst}
  \end{subfigure}\hfill
  \begin{subfigure}{0.32\textwidth}
    \centering
    \includegraphics[width=\linewidth]{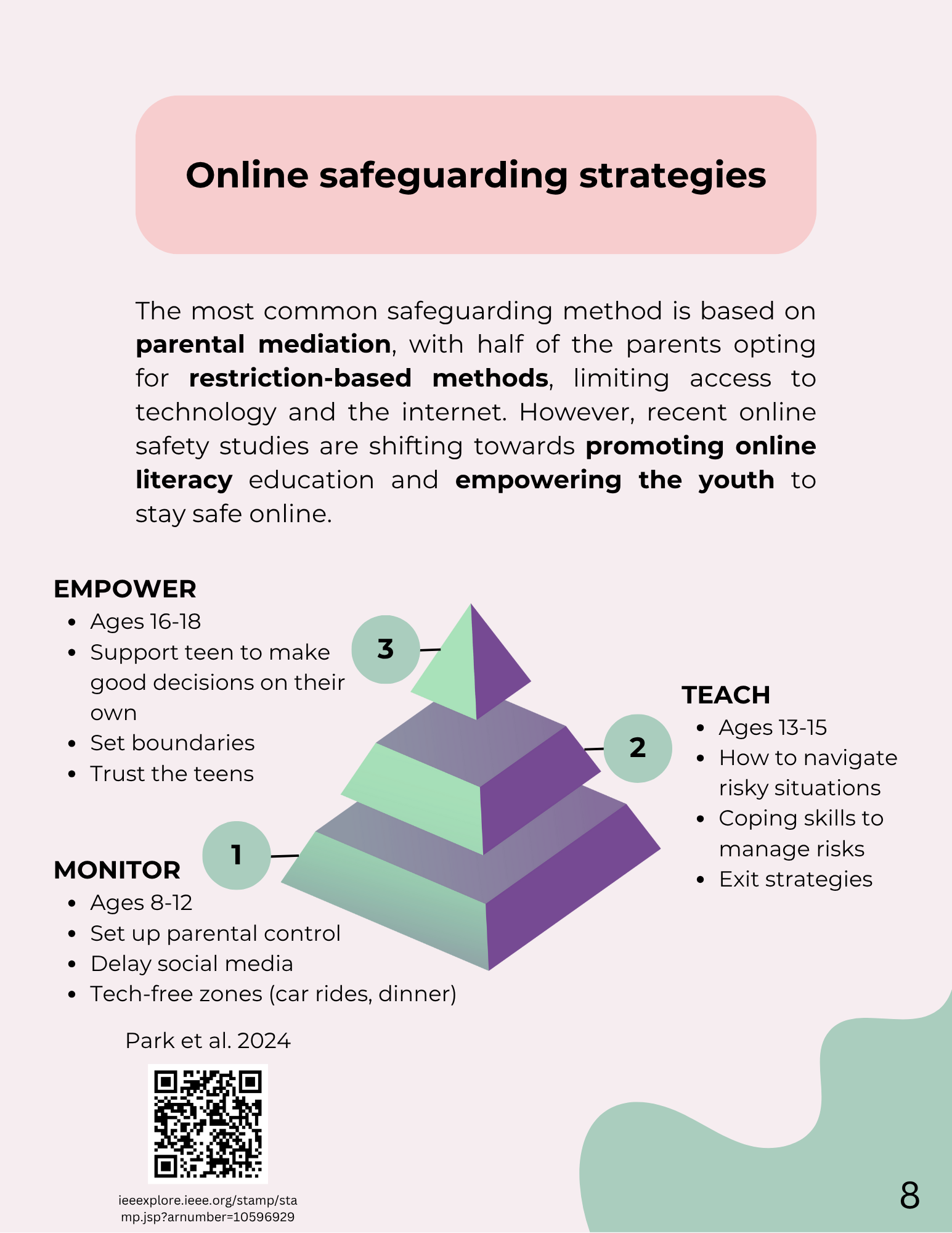}
    \label{fig:handbooksecond}
  \end{subfigure}\hfill
  \begin{subfigure}{0.32\textwidth}
    \centering
    \includegraphics[width=\linewidth]{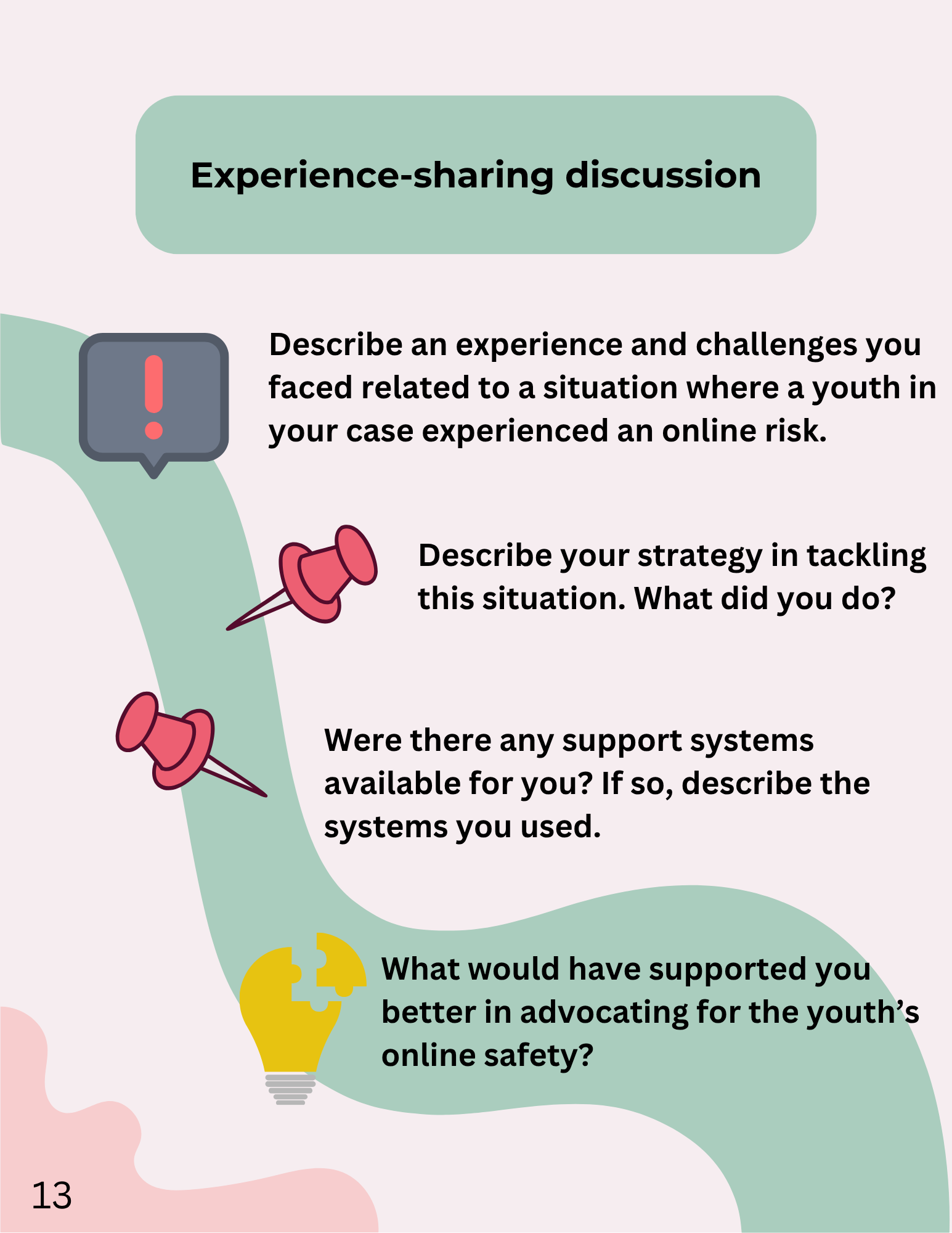}
    \label{fig:handbooksecond}
  \end{subfigure}

  \caption{Workshop handbook cover and sample pages}
  \label{fig:handbook}
\end{figure*}

We conducted an in-person workshop with (N=10) GALs in August 2024, lasting approximately three hours and comprising two parts: 1) content delivery and reflective discussion, and 2) a co-design activity. The workshop was audio-recorded, transcribed, and anonymized for analysis. Before beginning, participants reviewed and signed informed consent forms. Each participant received a handbook outlining the workshop’s objectives, summarizing key topics, and including resources on youth online safety (see \autoref{fig:handbook}).

\subsection{Part I: Content Delivery and Reflective Discussion}

An integral part of the workshop was to facilitate youth online safety training for GALS. Therefore, the first part of the workshop combined
a knowledge-sharing component with an open, reflective discussion to elicit participants’ perspectives and lived experiences advocating for the online safety of youth in the CWS. We opted for open large group discussions over smaller focus groups to foster collective dialogue and learning, allowing participants to build on each other’s experiences. This format enabled the exchange of diverse perspectives among GALs, who hold varying levels of experience and expertise, and encouraged shared reflection on systemic challenges and opportunities related to youth online safety. This portion of the workshop lasted approximately two hours. We began with participant introductions, where they shared their names, length of service, and motivations for becoming GALs. To prompt reflection, we asked an opening question: “\textit{How confident or well-prepared do you feel in advocating for the online safety of youth in the CWS?}” We then delivered a presentation summarizing current research and evidence-based safeguarding strategies. Throughout the presentation we guided an open discussion by prompts on topics such as challenges GALs face in supporting youth online safety, the support systems currently available to them, their observations of youth online behaviors, and the specific vulnerabilities of CWS-involved youth to online risks. This allowed participants to reflect on the information and relate it to their own practice.%We also delivered a presentation summarizing current research and evidence-based safeguarding strategies, embedded with open discussion in which participants reflected on the information and related it to their own practice.

\subsection{Part II: Co-Design Activity}

The second part of the workshop consisted of a one-hour co-design activity, in which participants were divided into two groups of five GALs, each facilitated by two researchers. The goal was to collaboratively envision solutions to promote the online safety of youth in the CWS. We adopted a co-design approach to engage GALs in the design process and elicit their situated expertise, consistent with participatory design principles that value stakeholders’ experiential knowledge \cite{kensing1998participatory}. We employed a Blue Sky Visioning approach \cite{gudowsky2017into}, encouraging participants to generate creative, aspirational ideas unconstrained by current technological or institutional limitations. Such an approach is particularly useful for complex social challenges, as it allows participants to articulate needs and possibilities beyond the constraints of existing systems. Participants were prompted with the following scenario:

\begin{quote}
    \textit{“You will now become technology designers. Think of a scenario of an online risk situation for a youth in the child welfare system, and based on this scenario ideate technology-based solutions. You can either redesign an existing technological system or come up with a completely new design that would promote the online safety of youth. The system can be preventative or interventional. Be creative!”}
\end{quote}

\begin{figure*}[!htbp]
  \centering
  \begin{subfigure}{0.31\textwidth}
    \centering
    \includegraphics[width=\linewidth]{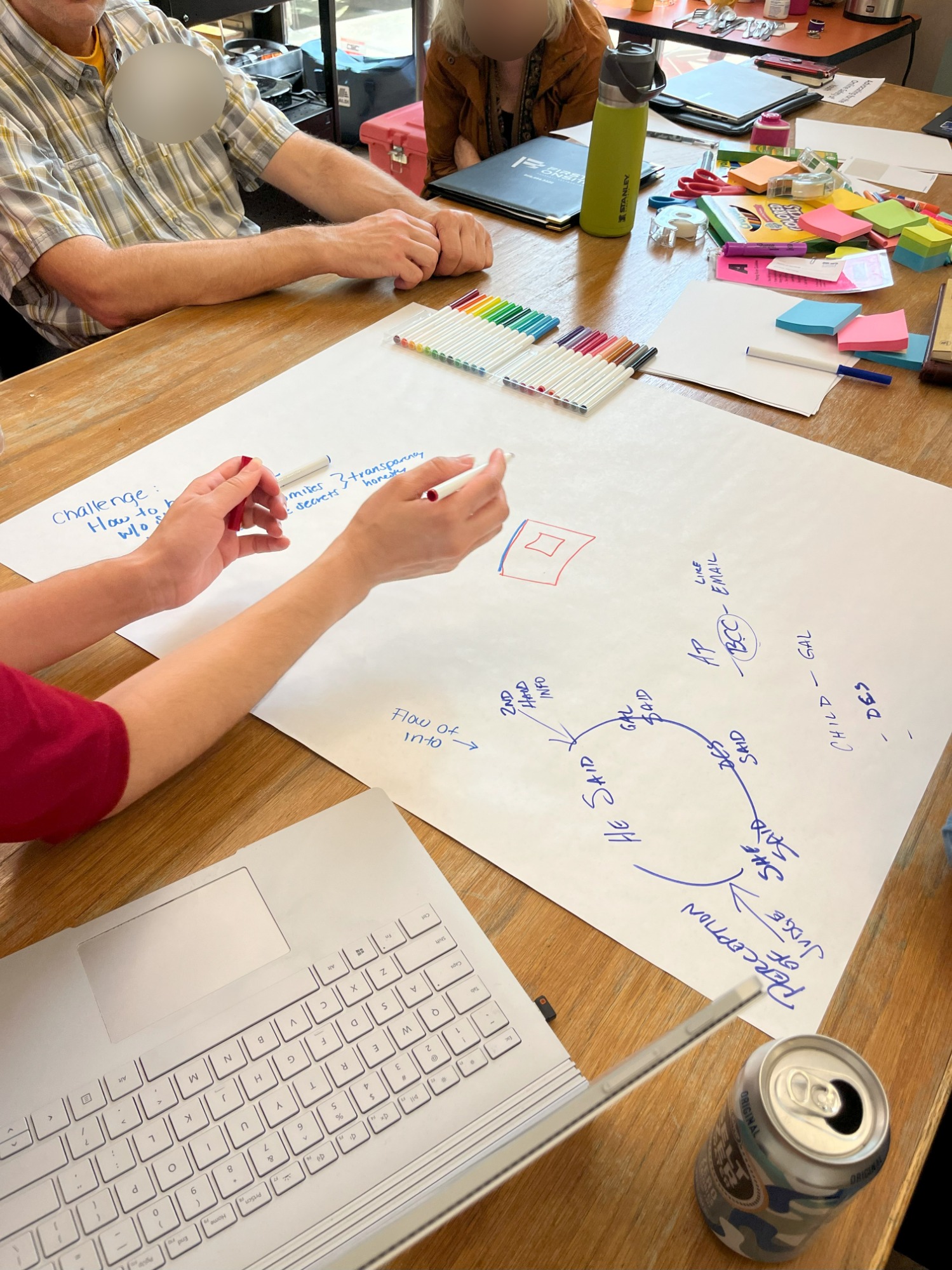}
    \label{fig:codesign1}
  \end{subfigure}\hfill
  \begin{subfigure}{0.31\textwidth}
    \centering
    \includegraphics[width=\linewidth]{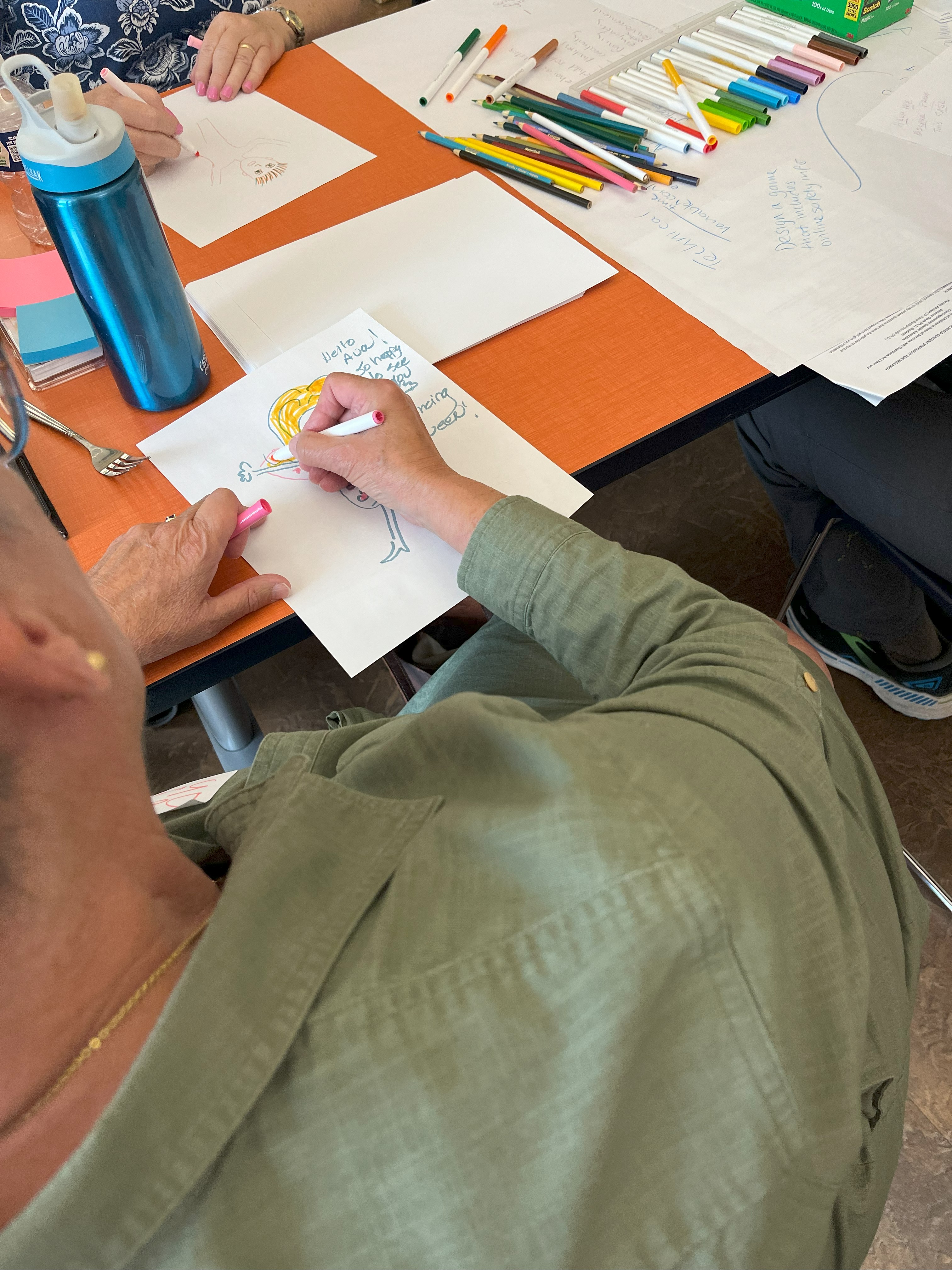}
    \label{fig:codesign2}
  \end{subfigure}\hfill
  \begin{subfigure}{0.31\textwidth}
    \centering
    \includegraphics[width=\linewidth]{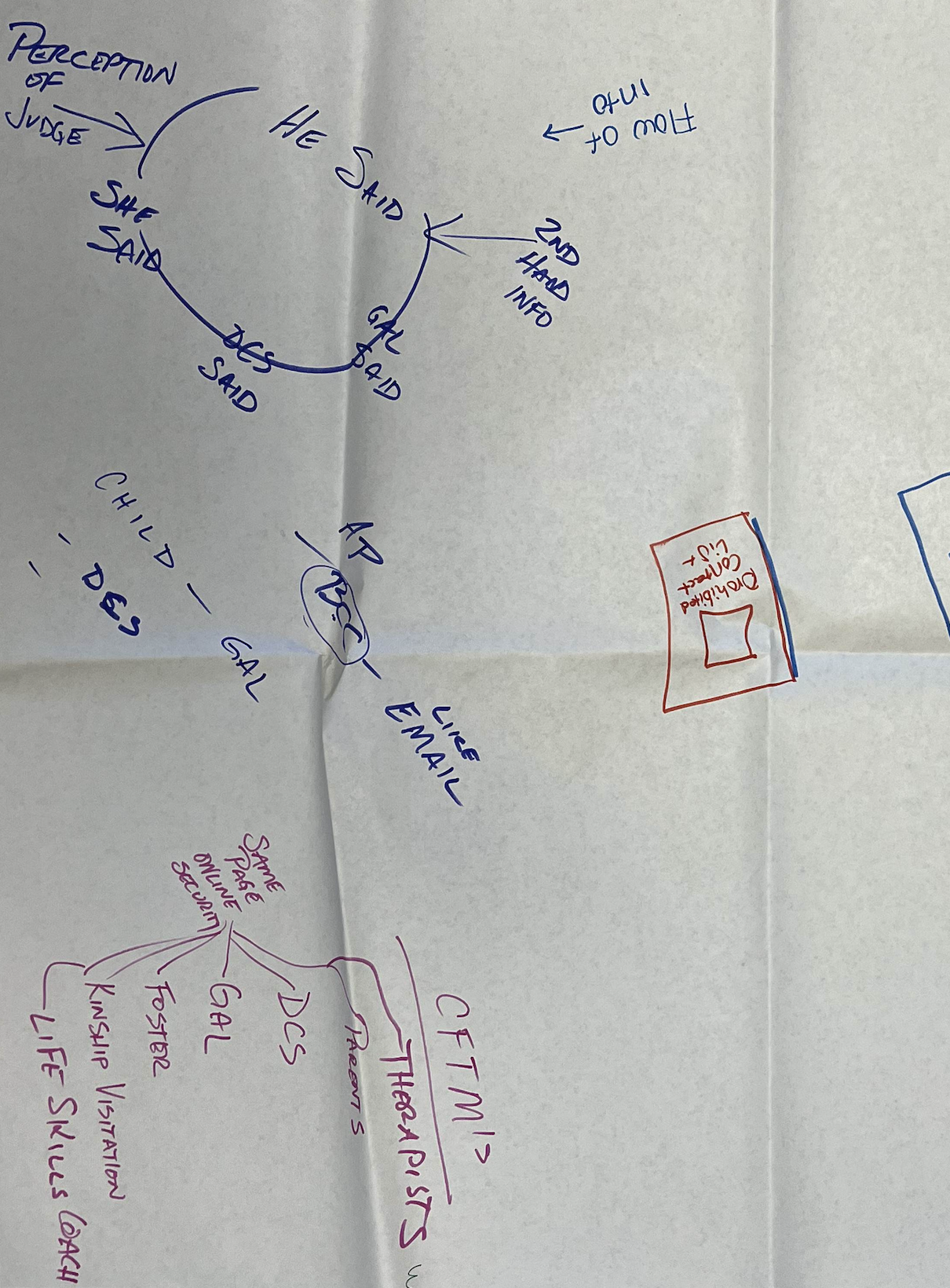}
    \label{fig:codesign3}
  \end{subfigure}

  \caption{Examples of participants brainstorming, sketching, and visualizing their ideated solutions to foster youth online safety.}
  \label{fig:codesign}
\end{figure*}

The session was structured with 45 minutes dedicated to small-group discussion and ideation, during which participants brainstormed, sketched, and wrote down their design ideas using paper, markers, and pens (see \autoref{fig:codesign}c). The participants visualized their concepts using mind maps and brainstorming features. In the final 15 minutes, each group presented their ideas to all participants, fostering collective reflection and peer feedback.

\subsection{Participant Recruitment and Demographics}

The study was approved by our university’s Institutional Review Board (IRB). Participant recruitment was conducted using a combination of snowball and convenience sampling over the course of one month (June-July 2024). We initially contacted a GAL program director in Indiana, who assisted with recruitment by directly reaching out to potential participants on our behalf. From there, we expanded the sample through participant referrals and additional outreach efforts. We sought practitioners who were 18 years of age or older and who were currently serving, or had served within the past five years, as a GAL/CASA to at least one teen (ages 13–17) in the U.S. child welfare system (CWS). Eligibility also required participants to speak English to ensure full participation in discussions and activities. By the end of July, we had recruited 10 participants, all of whom were active GALs at the time of the workshop. Although participants did not receive monetary compensation, they were offered the opportunity to engage with current research, reflect on their own practice, and share their experiences for dissemination to a wider audience, aligning with \citet{schon2017reflective} concept of reflection-on-action as a mode of stakeholder engagement. 

The participant group consisted of seven women and three men, aged between 53 and 78. Their experience as GALs ranged from seven months to 25 years. All participants identified as white and were predominantly retired professionals, with prior occupations including teacher, family physician, nurse practitioner, and academic roles. Given the sensitive nature of GALs’ work advocating for youth in the CWS and the potential for their comments to be traced back to specific cases or individuals, we took extra care to protect their privacy. Participants are anonymized throughout this paper to the fullest extent possible. \autoref{tab:participants} summarizes participant demographics, years of experience, and prior occupations.

\begin{table*}[tb]
\centering
    \caption{Participant Demographics and Background.}
    \label{tab:participants}
    \includegraphics[width=0.7\textwidth]{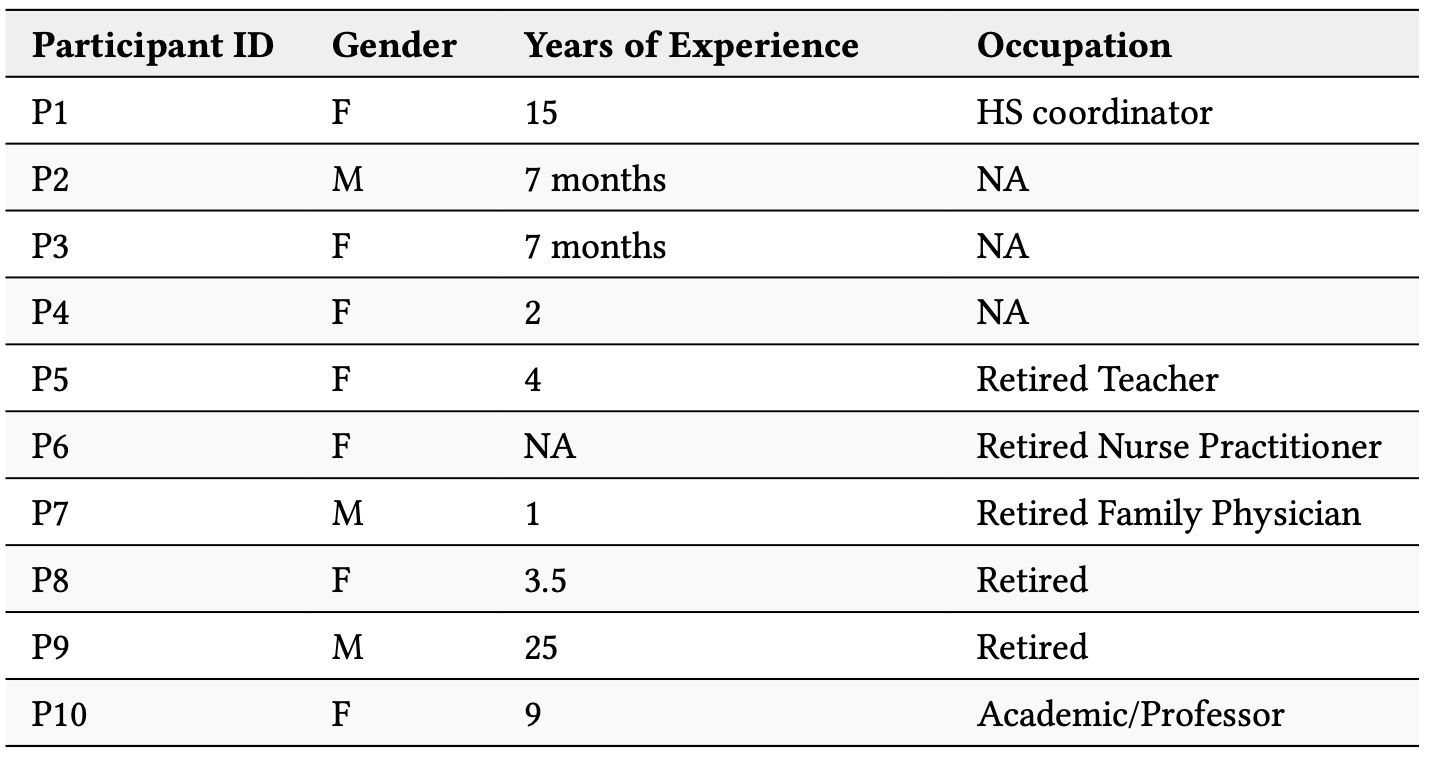}
\end{table*}

% \begin{table*}[tb]
% \scriptsize
% \caption{Participant Demographics and Background.}
% \label{tab:participants}
% {\renewcommand{\arraystretch}{1.5}
% \begin{tabular}{p{1.6cm}p{0.9cm}p{2.6cm}p{2.7cm}}
% \hline
% \rowcolor[HTML]{EFEFEF}
% \textbf{Participant ID} & \textbf{Gender} & \textbf{Years of Experience} & \textbf{Occupation} \\ \hline
% \rowcolor[HTML]{FFFFFF}
% P1 & F & 15 & HS coordinator \\ \hline
% \rowcolor[HTML]{F9F9F9}
% P2 & M & 7 months & NA\\ \hline
% \rowcolor[HTML]{FFFFFF}
% P3 & F & 7 months & NA \\ \hline
% \rowcolor[HTML]{F9F9F9}
% P4 & F & 2 & NA \\ \hline
% \rowcolor[HTML]{FFFFFF}
% P5 & F & 4 & Retired Teacher \\ \hline
% \rowcolor[HTML]{F9F9F9}
% P6 & F & NA & Retired Nurse Practitioner \\ \hline
% \rowcolor[HTML]{FFFFFF}
% P7 & M & 1 & Retired Family Physician \\ \hline
% \rowcolor[HTML]{F9F9F9}
% P8 & F & 3.5 & Retired \\ \hline
% \rowcolor[HTML]{FFFFFF}
% P9 & M & 25 & Retired  \\ \hline
% \rowcolor[HTML]{F9F9F9}
% P10 & F & 9 & Academic/Professor \\ \hline
% \end{tabular}}
% \end{table*}

\subsection{Data Analysis}

To answer our research questions, we conducted two separate reflexive thematic analyses \citet{braun2021thematic}. This method was chosen because it is well-suited to identifying patterns of meaning in qualitative data while acknowledging the active role of researchers in interpretation. An inductive coding process allowed themes to emerge directly from the data without being constrained by a predefined coding frame. We detail our analyses to follow.

\subsubsection{Analysis of Part I: Content Delivery and Reflective Discussion}

To understand what challenges families seek support for from GALs \textbf{(RQ1)} and the challenges GALs face when advocating for youth online safety \textbf{(RQ2)}, we conducted a thematic analysis on the transcripts from Part I of the workshop (reflective discussions). The first author first familiarized herself with the transcripts by carefully rereading the transcripts multiple times. The first author generated a set of initial codes for each RQ. Through an iterative process, the first author consulted with the last author to make sure codes appropriately reflected participants' discussions. After the coding process, the first and last author collaboratively generated themes to reflect the meaningful patterns in the data. Through iterative discussions the first, second, and last authors refined and mapped the themes and subthemes to our RQ1 and RQ2.

\subsubsection{Analysis of Part II: Co-Design Activity}

To examine how GALs envisioned technology-based interventions to support youth online safety \textbf{(RQ3)}, we conducted a thematic analysis on the transcripts and artifacts from Part II of the workshop (co-design activity). The second and third authors first familiarized themselves with the data by iteratively reviewing the transcripts and examining the design artifacts generated by participants. They independently developed initial codes for Group 1 and Group 2, regularly consulting with the first author to ensure consistency and clarity in the coding process. Through iterative discussions, the three authors refined and consolidated codes, resolving discrepancies through consensus and ensuring all codes appropriately reflected participants' ideas. The first author then examined the codes across both groups to identify overlapping design ideas and patterns in proposed features, grouping the codes into initial themes. Finally, the first, second, and last authors collaboratively refined the themes to reflect GALs’ envisioned design directions and priorities.
\section{Findings}

In this section, we present key findings organized by our research questions. We begin with RQ1, describing the challenges families face related to youth online safety according to GALs. We then address RQ2 by presenting the challenges GALs face in advocating for youth online safety. Finally, for RQ3, we describe the solutions GALs envisioned for effective online safety tools in the CWS and overarching themes that emerged from two co-design groups. We use illustrative quotes from participants to describe the themes that emerged from our qualitative data, where each quote is identified by the participant's ID, gender, and years of experience (if known). 

\subsection{Challenges Families Face Regarding Youth Online Safety According to GALs (RQ1)}

To understand the role GALs play in supporting families with online safety challenges, we examined the concerns related to online safety GALs observed or received as direct reports from families. These insights provide a foundation for understanding the specific issues GALs are asked to navigate and support. We identified three key challenges: (1) the dual nature of the internet, (2) youth lack digital literacy and online safety resources, and (3) intra-familial power dynamics. Below, we describe each theme in detail.

\subsubsection{The Dual Nature of the Internet}

GALs described internet access as both essential and unsafe for youth in the CWS, offering opportunities for connection, learning, and self-expression while also exposing youth to predatory, addictive, and high-risk interactions. According to GALs, this duality is deeply intertwined with the emotional and social challenges that many youth face, such as feelings of isolation, low self-esteem, and lack of trust in adults due to trauma and instability in their offline environment. To cope, they turn to online spaces \textit{"looking for some kind of affirmation and connection"} (P6, Female), feeling like these interactions are the only source of stability and validation, despite potential risks.

%GALs described internet access as both essential and unsafe for youth in the CWS, offering opportunities for connection, learning, and self-expression while also exposing youth to predatory, addictive, and high-risk interactions. According to GALs, this duality is deeply intertwined with the emotional and social challenges that many youth face\resubmit{, such as feelings of isolation, low self-esteem, and lack of trust in adults due to trauma and instability in their offline environment.} To cope, they turn to online spaces \textit{"looking for some kind of affirmation and connection"} (P6, Female). Several GALs emphasized that this search for validation online is often driven by fractured relationships with caregivers and frequent placement disruptions, leaving youth reluctant to \textit{"trusting that the parents are there for them"} (P3, Female, 7 mos. of experience). \resubmit{These emotional and social struggles can lead youth in the CWS to seek out online interactions that appear to meet their need for connection and validation.}

\begin{quote}
    \textit{"I think that they're just uniquely vulnerable to the adverse things online, a lot of them are struggling with self esteem and hard feeling connected and so they reach out to some of the riskier internet options because they can get that from them, although it may be high risk."}--P6, Female
\end{quote}

GALs observed that this reliance on digital spaces to fulfill unmet emotional needs often leads to risky behaviors, such as interacting with strangers, sexting, or becoming \textit{"totally addicted to playing online"} (P7, Male, 1 yr. of experience). Several participants noted that such patterns make youth easy targets for online predators who exploit their vulnerability and desire for connection.

%The reliance on digital spaces to fulfill unmet emotional needs leaves youth vulnerable to harm. GALs observed that many youth in the CWS engage in risky online behaviors, such as interacting with strangers, engaging in sexting, or becoming \textit{"totally addicted to playing online"} (P7, Male, 1 yr. of experience) in pursuit of connection or ways to cope with loneliness. Several GALs were concerned that the internet exposes youth to online predators who take advantage of their vulnerable emotional state and exploit their desire for connection.

\begin{quote}
    \textit{"They're reaching out to people that [they] probably shouldn't talk to. They're getting involved in sexting, you know, all kinds of stuff like that."}--(P5, Female, 4 yrs. of experience)
\end{quote}

On the other hand, several GALs acknowledged that online spaces can provide youth a sense of freedom and autonomy, allowing them to express themselves, connect with others who share similar experiences, and access information or guidance that might otherwise be unavailable to them, ultimately helping them build confidence, feel less alone, and gain a stronger sense of self.

\begin{quote}
    \textit{"They actually are looking the internet for answers."}--P4, Female, 2 yrs. of experience
\end{quote}

All in all, GALs recognized that the internet provides youth in the CWS with both risks and benefits to cope with their emotional and social needs. Families, however, often struggle to navigate this duality, unsure how to support youth in navigating digital environments that are simultaneously empowering and dangerous. This underscores the need for guidance and resources to help families and youth navigate digital spaces safely and meaningfully.

% \begin{quote}
%     \textit{"There's such a isolation element to them, that they will walk into something unaware of the red flags."}--P4, Female, 2 yrs. of experience
% \end{quote}

\subsubsection{Youth Lack Online Safety Resources and Support}

GALs reported that youth in the CWS often lack appropriate support and \textit{"resources that would help them navigate [online] positively"} (P3, Female, 7 mos. of experience). Without structured guidance, they are left to explore complex digital environments on their own while lacking the skills to recognize unsafe interactions or understand the implications of their actions online. Some GALs highlighted that these youth often face restricted access to technology and the internet due to court orders or rules set by foster parents or guardians. While such restrictions aim to protect them from harm, they can leave youth unprepared to manage risks when access is restored and further isolated from resources and opportunities for connection.

\begin{quote}
    \textit{"The problem with youth in residential care is that they are totally cut off. The funds are taken away from them. They're not supposed to have access to the Internet."}--P8, Female, 3.5 yrs. of experience
\end{quote}

Several GALs highlighted that youth in the CWS rarely have stable and trusting relationships with caregivers or advocates who support and guide them through their emotional challenges and online experiences. For instance, fractured relationships with caregivers and frequent placement disruptions leave youth reluctant to \textit{"trusting that the parents are there for them"} (P3, Female, 7 mos. of experience). Without trust and support, youth become easy targets for online predators who exploit this absence of oversight.

\begin{quote}
    \textit{"They're easily preyed upon because [...] they don't typically have the voice of a parent behind them."}--P4, Female, 2 yrs. of experience
\end{quote}

In the absence of guidance, many youth develop self-reliant and secretive coping strategies—\textit{"they're probably not going to tell you what's going on"} (P4, Female, 2 yrs. of experience). While this independence can foster resilience, it also obscures risky online behaviors and limit opportunities for intervention.

\begin{quote}
    \textit{"They're so used to solving problems on their own [...] and taking care of themselves and actually keeping themselves safe in ways that work in the physical world. I think sometimes they take that confidence with them into the digital space. [...] But I think sometimes [youth] can be overly confident in their ability to manage that [online] risk."}--P10, Female, 9 yrs. of experience
\end{quote}

This convergence of inadequate digital literacy resources, restrictive access policies, and absent supportive relationships leaves youth simultaneously denied opportunities to develop healthy online navigation skills while being expected to independently manage complex digital risks once they gain access. The resulting combination of inexperience, isolation, and overconfidence leaves these youth particularly ill-equipped to recognize and respond appropriately to online threats, highlighting the critical need for comprehensive digital safety support tailored to their unique circumstances.

\subsubsection{Intra-Familial Power Dynamics}

GALs described how complex and often strained intra-familial power dynamics in the CWS make it difficult for families to mediate youth’s online activities or establish consistent safety practices. These tensions often stem from placement instability, past trauma, and the use of technology as a form of control or leverage. Because many youth have experienced abandonment and unstable relationships with adults, they struggle to build trust and attachment, which in turn limits caregivers’ ability to provide guidance or enforce boundaries. GALs noted that these trust issues often surface in daily interactions, as youth emotionally withdraw and avoid sharing with adults out of fear of negative consequences.

\begin{quote}
    \textit{"[Caseworker] said that the child in our cases actually has shut down so that he has realized that by talking with her, there's been forces that have taken place in the case that he doesn't really want to have happen or he feels all negative and so he's just no answers, even if she has open-ended [questions], she's not going to get a whole lot from him."}--P5, Female, 4 yrs. of experience
\end{quote}

Beyond these relational challenges, GALs emphasized that frequent placement changes further compound the problem by disrupting continuity in care and technology supervision. Each new guardian often introduces different expectations and rules around technology use, leading to inconsistent mediation practices, where \textit{"one family's doing one thing and then the child gets put in another situation"} (P3, Female, 7 mos. of experience). GALs noted that these differences often swing between overly invasive monitoring and complete lack of oversight, leaving youth caught between extremes without stable guidance.

\begin{quote}
    \textit{"I have a teenage girl right now who has engaged in some online risky behaviors and one family she was with maybe were overly invasive, and the person she's with now may not be invasive enough."}--P6, Female
\end{quote}

In addition, GALs described how conflicting perspectives between biological and foster parents or guardians can further undermine caregivers’ authority and make it harder to enforce consistent boundaries around technology use. These tensions often create confusion for youth and complicate efforts to establish shared online safety practices. Furthermore, GALs noted that caregivers often struggle with weak authority over youth, as youth may resist or disregard rules and boundaries set around technology use.

% \begin{quote}
%     \textit{"If the [biological] parents are saying that [youth] should not have access to [technology devices], but foster parents [say] we need [youth] to have phone because we need to be able to be in touch with them."}--P4, Female, 2 yrs. of experience
% \end{quote}

Finally, GALs noted that technology itself often becomes entangled in power struggles between caregivers and youth. Several GALs highlighted that \textit{"there's this constant power dynamic going on"}--(P4, Female, 2 yrs. of experience), where caregivers use technology access as a "weapon" to gain leverage and control over youth or use it as a way to punish them. On the other hand of this power play, youth actively circumvent the restrictions and \textit{"will always find a way to work around"} (P8, Female, 3.5 yrs. of experience). Few GALs noted that biological parents may also use technology to gain leverage over the youth, mainly as means of manipulation or emotional abuse. Rather than fostering connection, these interactions further compromised youth safety and added complexity to already fragile family dynamics.

\begin{quote}
    \textit{"We found that several [biological] parents use texting as a weapon, abusive."}--P1, Female, 15 yrs. of experience
\end{quote}

% \begin{quote}
%     \textit{"I had another case where cell phones [were] being used as weapons between parent and child power struggle, like so I'm going to do this, I'll take away your phone."}--P10, Female, 9 yrs. of experience
% \end{quote}

These findings illustrate how fractured trust, inconsistent technology practices, and the use of technology in power plays creates a complex and vulnerable online environment for youth in the CWS. Such dynamics not only make it difficult for caregivers to enforce consistent online safety practices but also leave youth navigating digital spaces without stable guidance or support.

\subsection{Challenges GALs Face in Advocating for Youth Online Safety (RQ2)}

To contextualize the role of GALs in supporting youth online safety, we examined the specific challenges they face in advocating for youth in digital contexts. While GALs are positioned to act in the best interest of youth, they often operate within systems that limit their capacity to provide consistent, informed support. Our analysis revealed three key challenges GALs encounter in their advocacy efforts: (1) systemic barriers and limited institutional resources, (2) navigating communication across disconnected stakeholders, and (3) limited understanding of youth’s digital experiences and needs. %In this section, we elaborate on these challenges and their impact on GALs’ ability to support youth online safety.

\subsubsection{Systemic Barriers to Online Safety Advocacy}

GALs described how outdated court orders, case-specific inconsistencies, and limited institutional support hinder their ability to advocate for and adapt online safety measures as the needs of youth in the CWS evolve. In many cases, initial court restrictions on technology use no longer align with the youth’s developmental stage or current placement context, yet GALs struggle to update these directives in a timely or flexible way. Such overly broad and outdated restrictions create tensions between legal oversight and youth’s growing need for connection, autonomy, and digital inclusion.

\begin{quote}
    \textit{"I think one of the challenges for us [with] cases that drag on is getting full [court] orders to reflect that ongoing involving structure. You might have an initial order that's like zero contact, zero restriction or all restrictions or control. You know, we're trying to also then get the court system on board."}--P10, Female, 9 yrs. of experience
\end{quote}

GALs also pointed to case-specific inconsistencies, noting that the type of case they are assigned to, such as CHINS versus probation or custody, often determines their access to key resources like counseling services or digital safety guidance. As a result, some youth receive robust access to professional support while others do not.

\begin{quote}
    \textit{"Like if you're in CHINS case [it's] highly more likely that there's going to be counselors. But if you're working in probation case or custody case, those might not be available to you because you're outside of that system."}--P10, Female, 9 yrs. of experience
\end{quote}

%GALs emphasized limited access to formal online safety specific support, guidance, and resources that would aid them in effectively advocating for the digital safety of youth in the CWS. Many GALs reported that they had little direction beyond their own judgment, with no structured training or reliable point of contact to consult regarding concerns about youth technology use. Several GALs pointed out that their only source of information was the local GAL program director, stating that \textit{"we go to [GAL program director]"} (P8, Female, 3.5 yrs. of experience).

Moreover, GALs described challenges in tailoring their advocacy to youth of different ages and maturity levels, particularly when managing multiple cases simultaneously.

\begin{quote}
\textit{"One of the challenges is, you might have a case where you're with a three-year old, but then a case with a 16 year old [...] So what's appropriate for a 15 year old is not operating at a 5 year old level. [...] I think just the challenges around the age ranges is always the challenge."}--P10, Female, 9 yrs. of experience
\end{quote}

Finally, many GALs emphasized that they lacked formal training or clear institutional guidance for addressing youth technology use, relying instead on personal judgment or informal advice from program directors, stating that \textit{"we go to [local GAL program director]"} (P8, Female, 3.5 yrs. of experience). Without structured support, GALs felt underprepared to navigate complex online safety concerns.

% \begin{quote}
%     \textit{"In terms of official support, no."}--P1, Female, 15 yrs. of experience
% \end{quote}

% "it's a question of how trustworthy [the support] is." -P1

% In addition to inconsistent court orders and limited institutional support, GALs also noted the challenge of navigating inconsistent digital rules across cases due to frequent changes in placement families and professionals involved in the youth's care. This inconsistency makes it difficult for GALs to establish and maintain best practices for keeping youth safe online. 

% \begin{quote}
%     \textit{"I can tell you right now like a [biological] parent is going to have a different attitude than the foster family and I don't know who's right there."}--P4, Female, 2 yrs. of experience
% \end{quote}

%Finally, GALs are often assigned cases involving youth with widely varying ages and maturity levels, which complicates the implementation of age-appropriate strategies, especially when handling multiple cases simultaneously.

%Our findings indicate that systemic barriers constrain GALs’ ability to effectively advocate for youth online safety. Addressing these barriers requires not only updated policies and training but also a more flexible, youth-centered approach to accommodate the unique and evolving circumstances of each case.

Overall, these systemic barriers, such as outdated court orders, inconsistent case structures, and lack of institutional support, limit GALs’ ability to advocate for youth online safety. Addressing these challenges requires not only updated policies and targeted training, but also a more flexible, youth-centered approach that accounts for the diverse and evolving circumstances of each case.

\subsubsection{Navigating Complex Digital Relationships and Communication}
% \subsubsection{Navigating Disconnected and Unmonitored Communication Channels}
% \subsubsection{Communication Challenges Across Platforms and Stakeholders}
% \subsubsection{Balancing Monitoring, Privacy, and Fragmented Communication}
% \subsubsection{Monitoring Youth Communication in a Fragmented System}
% \subsubsection{Challenges in Overseeing Digital and Interpersonal Communication}

GALs described the fragmented and unstructured nature of digital communication between youth and the adults involved in their care as a major challenge to online safety advocacy. As part of their role, GALs must both foster and monitor youth–parent relationships, balancing the need for healthy connection with protection from potential harm. However, communication increasingly occurs across multiple, disconnected platforms, including those with disappearing messages (e.g., Snapchat), which complicates oversight and recordkeeping.

\begin{quote}
    \textit{"As a GAL, you're trying to foster that [youth-parent] relationship, but you also have to monitor that relationship. So something simple like, I had a parent and child interacting with Snapchat, well, everything goes away. There's nothing to record, that's not something that can be sent to me, right? So they have to screenshot that and send it to me because it'll disappear."}--P4, Female, 2 yrs. of experience
\end{quote}

For coordinated care, GALs often rely on tools like Our Family Wizard to facilitate and monitor youth-parent communication. Yet, many teens find ways to circumvent these controls. For instance, GALs noted that \textit{"if [youth] are denied a phone, they use a friends phone, and now you can't track that at all"} (P4, Female, 2 yrs. of experience), making consistent monitoring nearly impossible. GALs expressed that while oversight is essential for safety, they also recognized the importance of youth privacy and the delicate balance between protection and autonomy, noting the difficulty of navigating when and how to intervene.

\begin{quote}
    \textit{"The youths' need for privacy in certain circumstances but not in other circumstances. And you know how you navigate the online use by the youth without them feeling alienated, angry and having their privacy invaded? How do you kind of walk that advocacy?"}--P6, Female
\end{quote}

%This fragmented communication landscape complicates GALs’ efforts to provide consistent support while balancing the youth’s need for privacy, raising questions about how to advocate effectively without alienating or invading the youth’s sense of autonomy.

GALs further described challenges with fragmented communication across stakeholders, caseworkers, therapists, foster families, and biological parents, which often results in conflicting or secondhand information. They noted that information is often repeated, distorted, or outdated, making it difficult to discern what is accurate or actionable and to assess online risks effectively.

\begin{quote}
    \textit{"This flow of information has such a mess and I'm constantly trying to discern gossip, second hand and real information, because everything is easily received and reiterated. [...] We're just constantly trying to deal with what's true and what's not"}--P4, Female, 2 yrs. of experience
\end{quote}

Together, these challenges, such as ephemeral platforms, youth circumvention, privacy concerns, and fragmented communication among stakeholders, create a complex and often frustrating landscape for GALs to navigate. They must balance advocacy with oversight, protection with respect for autonomy, and incomplete information with the need to act decisively. Addressing these issues will require better tools, clearer protocols, and stronger cross-stakeholders collaboration to help GALs mediate youth’s digital relationships more effectively and equitably.

\subsubsection{Gaps in Understanding Youth Digital Experiences}

%- Intergenerational challenges --> intergenerational disconnect, GALs themselves do not understand youth online activities. They also need to navigate the disconnect between youth and parent intergenerational differences.

Most GALs expressed feeling uninformed when it comes to navigating the digital risks youth face, citing limited training, guidance, or access to expertise. Despite their role in advocating for youth wellbeing, GALs acknowledged substantial gaps in their understanding of common platforms, emerging technologies, and online threats such as AI-driven content, geofencing, phishing, and online sexual abuse. For many, this lack of knowledge undermined their confidence in addressing online safety concerns.

\begin{quote}
    \textit{"Ignorance is my reason. [...] [I'm] not confident because I don't know enough."}--P7, Male, 1 yr. of experience
\end{quote}

Even those who reported greater awareness of online safety felt overwhelmed with the rapidly evolving online landscape. The sense of constantly falling behind left them frustrated and uncertain about how to best support youth.

\begin{quote}
    \textit{"I went for [not confident] and a lot of that for me is these changes are so fast, and if you're just not constantly in the know or on top of it."}--P8, Female, 3.5 yrs. of experience
\end{quote}

All GALs in our study were over 50 years old, and many reflected on how their generational distance from youth made it difficult to relate to or fully grasp the appeal and risks of youth’s online activities. They described their own limited familiarity with gaming, social media culture, and online norms as a barrier to connecting with youth or being able to provide appropriate guidance.

\begin{quote}
    \textit{"I didn't grow up playing with the computer. I use it when I need to. I can make use of it. But I don't play games at all with the computer. So because of that, I don't understand what kids are going through and the draw of computer games."}--P7, Male, 1 yr. of experience
\end{quote}

These intergenerational disconnects extended beyond GAL–youth relationships to those between youth and their caregivers. GALs often had to mediate when caregivers’ rules and expectations, which were shaped by their own limited digital understanding, clashed with youth’s online behaviors. Such dynamics further complicated efforts to establish consistent and realistic online safety practices.

\subsection{Online Safety Solutions Designed by GALs for the CWS (RQ3)}

We co-designed with GALs technology-based solutions that could better address the online safety needs of youth in the CWS. Drawing from their unique position within the CWS, GALs emphasized the importance of systems that go beyond restriction and surveillance to promote trust, stability, and connection. GALs envisioned solutions that strengthen youth agency, enable better communication across stakeholders, and offer therapeutic and educational support.

Group 1 focused on the underlying issues of trust and attachment formation of youth in the CWS and their mental health disorders, such as Reactive Attachment Disorder (RAD), which impact their online activities. They envisioned an avatar-based support system designed to help youth in the CWS develop healthier digital habits, emotional regulation, and trust. The avatar would act as a companion and therapeutic guide, offering positive reinforcement, modeling empathy, and integrating trauma-informed interventions like cognitive behavioral therapy (CBT)-inspired activities. Customizable in appearance and personality, the avatar would promote self-expression and relational growth while gently encouraging offline engagement. To balance autonomy and safety, GALs proposed tiered oversight mechanisms that preserve youth privacy while alerting trusted adults only when serious risks arise. Overall, the concept reframes online safety as relational empowerment rather than restriction.

Group 2 focused on fragmented communication among stakeholders as a fundamental barrier to youth online safety. To tackle this, they envisioned a secure, unified communication platform to bridge the fragmented dialogue among stakeholders in the CWS. Designed to facilitate transparent, court-approved, and trauma-informed communication, the system would feature tiered access controls, real-time supervision, and persistent conversation channels that support early and continuous engagement around digital risks. Centered on youth trust and empowerment, it would give young people agency over communication boundaries while ensuring accountability among adults. By streamlining information flow and embedding online safety into routine casework, the proposed platform reframes digital engagement in the CWS as collaborative, preventative, and emotionally safe rather than reactive or punitive.

Below, we disseminate the overarching themes identified from the analysis of the two design ideas proposed by GALs. Note that quotes are attributed to either Group 1 or Group 2 rather than individual GALs.

%Below, we first describe the individual solutions that each of the two groups of GALs developed, followed by a thematic synthesis of the design principles that emerged across both groups. 

\subsubsection{Fostering Coordinated Stakeholder Communication with Safety Measures}

GALs emphasized the need for a unified, secure platform connecting all key CWS stakeholders, including youth, GALs, caseworkers, therapists, and caregivers, to address the fragmented communication that undermines youth online safety. GALs noted that \textit{"there's so many players that you're constantly talking to"}--Group 2 (P4, Female, 2 yrs. of experience), making it difficult to discern the true information. To reduce this misinformation and ensure critical behavioral cues are not missed, GALs proposed leveraging a platform with smaller channels for targeted conversations, allowing stakeholders to share important perspectives more effectively.

\begin{quote}
    \textit{"A lot of people never said what they wanted to say at the meeting, so this would be an opportunity to do that when you'd have smaller channels. Like the GALs and the child or the GALS and the parent, [...] you know those sort of individual conversations."}--Group 2 (P1, Female, 15 yrs. of experience)
\end{quote}

Coordinated communication would also address systemic supervision challenges that hinder timely conversations on youth's wellbeing, including online safety, noting that \textit{"the ability to check in even for a minute or two minutes is hugely important"} (Group 2, P1, Female, 15 yrs. of experience) when it comes to youth's wellbeing. GALs further noted that relying on scheduled third-party supervisors often delays or prevents parent-youth interactions. They proposed features such as real-time supervision, communication logs, and recorded threads that could be activated as needed for sensitive interactions, ensuring safe digital engagement even when in-person supervision is not possible.

\begin{quote}
\textit{"There's an excuse that I feel like keeps getting used is we couldn't schedule, whereas if this [feature] was set up that there's somebody watching this conversation [real-time], [youth and parent] can still have the conversation. They can't use the excuse that there’s no supervisor available."}--Group 2 (P4, Female, 2 yrs. of experience)
\end{quote}

Most importantly for online safety, GALs noted that a unified platform would enable earlier and continuous engagement around digital risks and boundaries. Group 2 emphasized that online safety conversations often begin too late, typically around age 16, underscoring the need for earlier education. The platform would facilitate ongoing, structured dialogue among stakeholders, helping youth develop healthy online behaviors and ensuring preventive measures are implemented before harm occurs.

\begin{quote}
    \textit{"The app could help standardize information-sharing and support more consistent messaging around online security and other issues. [...] Ideally, online safety conversations should happen earlier than when collaborative care typically starts at age 16."}--Group 2 (P4, Female, 2 yrs. of experience)
\end{quote}

To balance oversight with confidentiality, GALs from both groups highlighted the importance of tiered oversight of youth's online communication and activities. For instance, Group 2 proposed communication platform was visioned to feature court-approved messaging capabilities and tiered access controls, allowing stakeholders, like GALs, therapists, caseworkers, and parents, to have different levels of visibility depending on their role and the youth's specific circumstances.

\begin{quote}
    \textit{"That you have this [BCC] capacity so that there's this secure exchange between let's just say child and GAL. But there's also BCC capacity so that I could add on [caseworker], you know, so she's watching the conversation, too. And we may not want that, but I'm saying like, the capability that you could add someone to that specific correspondence."}--Group 2 (P4, Female, 2 yrs. of experience)
\end{quote}

Similarly, Group 1 emphasized tiered oversight for their proposed avatar system, providing different levels of access to sensitive information flagged by built-in firewalls that detect concerning behaviors, such as expressions of self-harm or aggression. Firewalls would activate only when the avatar's supportive interventions fail, automatically alerting designated adults via text, phone, or email. This approach ensures oversight is used as a last resort, protecting youth privacy. While GALs noted that \textit{"engaging the parent or the foster [parent] as much as they can, [...] can be beneficial to both the child and the foster parent"} (Group 1, P7, Male, 1 yr. of experience), access to sensitive alerts would be restricted, as GALs observed that \textit{“with some of the parents, I wouldn’t necessarily want that information going to them”} (Group 1, P10, Female, 9 yrs. of experience).

\begin{quote}
    \textit{"There has to be some things that if that person's going to hurt somebody or hurt themselves that, you know, somebody has to be notified. [...] On the other hand, there has to be a degree of confidentiality to it."}--Group 1 (P9, Male, 25 yrs. of experience)
\end{quote}

Overall, GALs envisioned coordinated communication systems that balance timely oversight with youth confidentiality, using secure platforms, tiered access, and real-time monitoring features. By reducing fragmentation, promoting transparency, and ensuring appropriate intervention when necessary, these systems aim to support both youth wellbeing and effective collaboration among all stakeholders in the CWS.

% \begin{quote}
%     \textit{"Is there a way to manage information so it's cleaner, cause it feels like it's just constantly muddy [...] I mean, it's just communication, right? Like you have to work at it, but [...] how do we use that medium that we already have [CFTM] to get everybody on the same page with online security and who's moving that ball down the the field?"}--Group 2 (P4, Female, 2 yrs. of experience)
% \end{quote}

%\paragraph{Tiered Oversight And Automated Alerts/Firewalls}

\subsubsection{Promoting Trust-Building and Emotional Safety Through Technology}

A central theme across both groups was the importance of reinforcing emotional safety, empowerment, and trust through technology. GALs described that youth in the CWS often struggle with trust and attachment, and may be diagnosed with Reactive Attachment Disorder (RAD) due to early neglect or trauma, which impairs their ability to form healthy emotional bonds. As a result, youth are often hesitant to share information due to past experiences with broken trust or fear of consequences.

\begin{quote}
    \textit{"It has to be a dialogue, otherwise it becomes a power play [...] It is all how to make dialogue effective, without it being manipulated. So [youth] know it's not going to be used against them."}--Group 2 (P3, Female, 7 mos. of experience)
\end{quote}

The envisioned avatar system by Group 1 was primarily designed to support youth by promoting safe online behavior, emotional regulation, and relational development. Being integrated as a system overlay across the youth’s online environments, the avatar would be serving as a "buddy" that builds rapport over time while offering consistent, non-judgmental support, GALs noting that \textit{"[youth] are building trust with their avatar"} (Group 1, P8, Female, 3.5 yrs. of experience). GALs also emphasized promoting youth self-expression through designing their own avatar to further facilitate trust-building. The avatar could take any form, from animate to inanimate objects, and have special capabilities such as turning back time or flying. While youth may be able to create a monster, the avatar's socio-emotional traits \textit{"would all be positive"} (Group 1, P6, Female), embodying kindness, empathy, compassion, humor, and creativity.

\begin{quote}
    \textit{"Your avatar can just be a substitute to human being with all the characteristics that we want all human beings to have."}--Group 1 (P8, Female, 3.5 yrs. of experience)
\end{quote}

GALs in Group 2 further built on the theme that youth in the CWS are often hesitant to share information due to past experiences with broken trust or fear of consequences.  Group 2 emphasized that effective online safety conversations require youth to feel emotionally safe and in control of their communications, without fear that their words will be used against them. Their envisioned unified communication platform would support this by maintaining transparency about who has access to conversations while giving youth agency in determining communication boundaries, reinforcing trust and promoting open dialogue across stakeholders.

\begin{quote}
    \textit{"If this is designed to build trust between you and the child then [...] you got to be honest about it."}--Group 2 (P1, Female, 15 yrs. of experience)
\end{quote}

Beyond trust-building features, GALs in Group 1 envisioned integrating therapeutic capabilities directly into the avatar system by embedding evidence-based interventions, such as cognitive behavioral therapy (CBT) inspired activities and calming break spaces, into gaming experience to support youth in regulating emotions and processing trauma. During natural pauses in gameplay, youth could enter immersive environments like gardens or forests that use growth and renewal metaphors to help them recognize the transitory nature of difficult emotions. Additionally, the avatar could be programmed with trauma-informed principles to provide AI-driven CBT support, facilitate talk therapy, enable youth to create and save stories for personalized therapeutic engagement, and guide them through safely revisiting traumatic events. By integrating these features, the avatar system would extend its trust-building role to provide accessible, emotionally safe, and engaging mental health support for youth in the CWS.

\begin{quote}
    \textit{"The game could be designed in a way that it's cognizant of trauma-focused therapy principles or approaches, and some of those principles were built into the game. [...] Ultimately, if we could use an AI therapist, would be perfect."}--Group 1 (P7, Male, 1 yr. of experience)
\end{quote}

Overall, incorporating features such as transparency, trust-building, and trauma-informed interventions would promote youth agency while providing a supportive framework of care. GALs emphasized that technology should not control or restrict youth, but rather be designed with empathy to foster open, trusting dialogue. In this vision, online safety becomes a form of relational support that empowers youth to navigate digital spaces with confidence, balance, and emotional security.

\subsubsection{Supporting Real-World Relationships and Self-Regulation Through Technology}

GALs across both co-design groups emphasized that technologies should inherently support, not replace, real-world relationships and self-regulation. GALs envisioned built-in features, such as timers, countdown notifications, and automatic shut-offs, that would help youth gradually disengage from online activities, supporting healthier digital habits. These tools were envisioned as universal safety mechanisms embedded across digital platforms. Beyond limiting screen time, GALs also suggested integrating constructive alternatives that would be educational, guide youth towards safe online practices, and foster social support to promote self-regulation.

\begin{quote}
    \textit{"What are some of the alternatives to gaming that we could get this youth interested in, or alternative games? You know, that might be more instructive and help their development."}--Group 1 (P7, Male, 1 yr. of experience)
\end{quote}

Group 1 emphasized that the avatar should gently guide youth toward healthier digital habits. Drawing inspiration from smart technologies that remind users to engage in self-care (e.g., standing up, drinking water), GALs envisioned the avatar providing timely prompts for youth to take breaks from gaming and transition into other activities, such as educational games or offline activities. These prompts would be delivered in a supportive, non-punitive tone and reinforced through encouraging messages to help youth gradually build safer and more balanced online routines.

\begin{quote}
    \textit{"A positive reinforcement of you've accomplished something today. You played that game for a while, and now you're shifting gears for a few minutes and it's just awesome."}--Group 1 (P6, Female)
\end{quote}

GALs in Group 2 highlighted that promoting online safety requires balancing digital engagement with healthy offline activities. They proposed incorporating built-in curfews and certain limits within the communication platform, noting that \textit{"no child should be online at 2AM"} (Group 2, P4, Female, 2 yrs. of experience). GALs suggested that the platform would encourage activities like reading print books and offline journaling as essential components of building focus, emotional resilience, and healthy relationships with technology.

\begin{quote}
    \textit{"I'll be honest, if you want to talk about online stuff I would like kids to read more books."}--Group 2 (P4, Female, 2 yrs. of experience)
\end{quote}

GALs especially highlighted the importance of encouraging offline engagement to support real-world relational development. Given that youth in the CWS often struggle to form and maintain relationships, the systems should be designed to facilitate gradual online-to-offline transitions by prompting collaborative or social activities and \textit{"encouraging actual human interaction"} (Group 1, P7, Male, 1 yr. of experience). For example, Group 1 noted that the avatar might encourage youth to spend time with a caregiver or to join peers in a shared activity like geocaching, helping build relationships at the youth’s own pace. Additionally, recognizing that some youth with RAD may struggle to understand and express empathy, GALs suggested that the avatar could model emotional insight through interactive storylines, helping youth identify and respond to others’ feelings in a supportive, guided manner.

\begin{quote}
    \textit{"One other things that is a characteristic [to] RAD, is the inability to experience empathy. So I think we need to build in the game some way of not only encouraging but modeling how to sense how others are feeling and how to express that."}--Group 1 (P7, Male, 1 yr. of experience)
\end{quote}

Rather than relying solely on monitoring or blocking, GALs recommended online safety tools that promote self-regulation and encourage youth towards healthy transitions away from screens and into restorative offline activities. Features that promote offline activities and social connection, such as journaling, reading, or shared experiences with trusted adults and peers, were seen as especially valuable. This way, technology becomes a relational bridge, guiding youth between digital and physical spaces while supporting emotional growth and safe, grounded interactions.

Overall, GALs emphasized that digital safety tools for youth in the CWS must extend beyond digital restrictive measures and surveillance. Instead, they should be integrated into a broader ecosystem of support that fosters trust, supports stakeholder collaboration, and bridges digital tools with offline development to ensure a holistic approach to youth online safety. Ultimately, this vision reimagines online safety not as restriction or control, but as a form of relational support that empowers youth to navigate digital spaces with greater confidence, balance, and emotional security.

\section{Discussion}

Our study explored how GALs perceive and advocate for the online safety of youth in the CWS, the challenges they encounter, and the technology-based solutions they envision. In this section, we reflect on our findings to articulate two key insights: the need to reconceptualize online safety as a value rooted in relational care, and the importance of treating it as a shared, multi-stakeholder responsibility. We also discuss the implications for socio-technical design, emphasizing trust-building, coordination, and youth empowerment, and conclude with limitations and directions for future research.

%\subsubsection{Integrating Technological Needs for Holistic Online Safety}

\subsection{Reconceptualizing Online Safety as a Value of Relational Care}

Youth in the CWS have often experienced significant relational trauma, including abandonment, maltreatment, and disruption of foundational relationships with birth parents, siblings, and friends. These experiences leave youth without stable, meaningful bonds, compromising their sense of belonging, trust, and safety \cite{beyerlein2014need, zhang2021trauma}. Prior studies have shown that such trauma affects not only emotional development but also how youth navigate social contexts, both offline and online \cite{oguine2025Chins, Badillo-Urquiola2017abandoned, gustavsson2015positive, Tanni2024May}. As GALs in our study noted, the internet often becomes a space where these youth seek the affirmation and connection that remain unmet in their offline lives. From this perspective, online behaviors often labeled risky can be reframed as adaptive responses to relational deprivation, a means of fulfilling the fundamental human need for connection. This trauma also undermines youth's ability to trust or accept adult guidance due to past experience of abandonment and betrayal by caregivers. Consequently, those most in need of supportive relationships are often least able to access them, leaving youth to navigate complex online environments without adequate support, as documented in prior studies \cite{caddle2022challenge, badillo2017understanding, sage2022systematic, kumar2025cultivating, oguine2023you}.

Despite these complex relational needs, our findings show that current online safety approaches in the CWS remain dominated by restriction, surveillance, and court-ordered digital limitations, echoing prior research \cite{badillo2019risk, badillo2024caseworker, kumar2025cultivating}. Such strategies misalign with the relational nature of safety as they overlook youth’s underlying need for connection and instead reinforce patterns of control and disempowerment that mirror prior trauma \cite{badillo2019risk, oguine2025Chins, zhang2025dangerous}. Our findings further reveal that youth often circumvent these restrictions, engaging online in secret and without support. Based on these insights, we propose that effective online safety for youth in the CWS must be grounded in relational care rather than behavioral control. Relational care approach align with established evidence demonstrating the effectiveness of relational care approaches \cite{alford2019use, denby2016promoting}, emphasizing trust-building, emotional connection, and collaborative support, have proven particularly effective for youth who have experienced betrayal trauma \cite{gomez2016shifting}. Trust-Based Relational Intervention (TBRI), for instance, provides a framework for addressing complex developmental trauma through connection, empowerment, and emotional regulation rather than behavioral compliance \cite{purvis2013trust}. Studies have likewise demonstrated that agency, genuine support, and emotional connection promote relational permanency and positive outcomes in foster care \cite{ball2021agency}, helping youth navigate transitions and challenges within the care system \cite{munford2022children}. Within online safety contexts, this requires reframing our focus from "How do we stop youth from engaging in risky online behaviors?" to "How do we build trust-based relationships that enable youth to seek support when they encounter online challenges?" This is supported by our findings, where GALs prioritized trust-building, emotional support, and collaboration over restriction and surveillance, echoing prior HCI scholarship that emphasizes the importance of designing for open communication and trust between youth and adults in digital safety contexts \cite{rutkowski2021family, ghosh2020circle, zhang2025dangerous, badillo2019stranger}. Reconceptualizing online safety as a value of relational care means recognizing that digital wellbeing is inseparable from emotional and relational health and youth in the CWS need trustworthy adults who offer guidance without judgment and control. This approach centers youth’s emotional wellbeing, agency, and need for secure, trusting relationships that support them across online and offline contexts.

\subsection{Online Safety as a Multi-Stakeholder Responsibility}

Unlike typical youth, youth in the CWS navigate their online experiences within a complex care network involving GALs, caseworkers, therapists, foster parents or guardians, biological parents, and other stakeholders. Consequently, youth often lack stable environments that consistently guide and support them through online challenges. While prior research has emphasized that online safety is inherently a joint responsibility that must extend beyond individual caregivers \cite{caddle2025building, akter2022parental, mark2017invitation, oguine2025multistakeholder}, this shared responsibility becomes more challenging in the CWS, where care teams are not only diffuse but often operate in silos, making online safety responsibilities fragmented and inconsistently enacted across cases. Prior research has similarly highlighted systemic fragmentation within the CWS \cite{lee2015accessing, edwards2023administrative, akin2017successes}, characterized by siloed information flows and disparities in accessibility among stakeholders such as foster parents, clinicians, and social workers \cite{hwang2017information}. GALs in our study echoed these challenges, describing a lack of online safety–specific guidance and limited coordination across care teams, consistent with prior findings from interviews with foster parents \cite{badillo2019risk} and caseworkers \cite{badillo2024caseworker}. These communication barriers leave youth vulnerable to online risks and often result in reactive, restrictive responses rather than proactive engagement with their digital experiences \cite{oguine2024mainstream}. The concept of “secondary stakeholders” in youth online safety \cite{qadir2024towards}, i.e., adults who influence youth safety but are not primary guardians, is particularly relevant here: in child welfare, this network expands to include legal advocates, therapists, and others, making collective responsibility for online safety even more difficult to achieve.

While many scholars have noted the importance of clearly defined roles among stakeholders in the CWS \cite{chuang2010role, darlington2005practice, leathers2009context} as well as structured collaboration and communication \cite{marie2025foster}, our findings indicate that limited coordination across care teams continues to pose challenges in effectively addressing youth online safety. In response, we advocate for reframing online safety in the CWS as a distributed, multi-stakeholder responsibility. Rather than overburdening individual actors, digital safety must be designed as a shared effort supported by clear role definitions, coordinated communication, and robust infrastructure, allowing for more consistent and comprehensive efforts to support youth in their online experiences while reducing the ambiguity of oversight. This aligns with Caddle et al.'s \cite{caddle2025building} framework for multi-stakeholder digital safety, which emphasizes the need for coordination over fragmentation when it comes to youth protection online. Ultimately, safeguarding youth online in the CWS requires infrastructural supports that facilitate collective and consistent care, which have shown improved outcomes in child welfare contexts \cite{glisson1998effects, smith2022collective}. Communication technologies, when designed as infrastructural supports rather than supervisory tools, can sustain coordination across dynamic care teams and ensure that no youth falls through the cracks due to unclear responsibility. %Online safety should be no different, it must be shared, structured, and embedded in the daily routines of a connected care team.

\subsection{Implications for Socio-Technical Design}

Our findings point to a critical need for reconceptualizing the design of online safety technologies for youth in the CWS. Systems must move beyond traditional models of restriction and surveillance to foster trust, coordinated care, and empower youth agency. Below, we outline key design implications for systems that aim to support online safety specifically for youth in the CWS.

\subsubsection{Designing for Multi-Stakeholder Collaboration}

Given the distributed and dynamic nature of care in the CWS, digital systems must facilitate coordinated communication across all stakeholders, including GALs, caseworkers, youth, and caregivers. Our findings show that centralizing communication is critical for enabling proactive and consistent online safety support for youth. Systems should support case-specific threads, enable role-based visibility of conversations with appropriate confidentiality settings, and incorporate context-aware messaging that flags sensitive or urgent content related to digital wellbeing concerns. When youth report online risks, automated alerts can ensure each stakeholder receives role-appropriate information, as highlighted by GALs to protect youth's privacy. Furthermore, multi-perspective risk analysis tools should aggregate online safety observations from different stakeholders to create comprehensive risk profiles that can collectively identify emerging online safety issues before they escalate \cite{multistakeholder2025}. By providing lightweight, role-aware infrastructures, these systems would facilitate timely decision-making around digital wellbeing through comprehensive safety planning workspaces \cite{agha2022realtime}. Moreover, such communication platforms should include built-in prompt templates that guide different stakeholders to initiate online safety discussions appropriately for their roles, as well as shared libraries of digital safety resources to ensure consistent messaging across the care team without overwhelming youth with duplicate information \cite{oguine2025Chins}. Ultimately, rather than relying on surveillance or control, these systems should promote empathetic guidance and affirming dialogue between youth and caregivers or advocates, incorporating youth-controlled disclosure mechanisms that allow young people to selectively share online safety concerns with specific stakeholders alongside clear explanations of mandatory reporting requirements \cite{oguine2025Chins}, contributing to transparent communication while protecting the privacy of youth that GALs emphasized as essential.

\subsubsection{Designing for Trust-Based Communication}

Online safety technologies should be based on relational care principles by creating interfaces that facilitate gradual trust-building through consistent, supportive interactions. This focus on trust was emphasized by GALs in our study, who highlighted the importance of fostering open, transparent communication between youth and their care networks. Technology design must therefore prioritize transparency and youth voice, aiming to strengthen relational bonds rather than enforce control. Previous research has shown that fostering trust between youth and caregivers promotes timely conversations about potential online risks, facilitating more effective online safety \cite{ghosh2020circle, cortesi2025frontiers}. Designing for trust means centering user experience around connection, affirmation, and emotional validation, principles that are especially vital for youth in the CWS. Systems should allow youth to manage their privacy by selectively sharing sensitive information with trusted adults while making non-sensitive updates visible to the broader care team, as noted by GALs. At the same time, technologies must support mandatory reporting of high-risk situations (e.g., suicide ideation) while maintaining confidentiality for everyday conversations. Transparency between youth and the care team could be enhanced through real-time notifications showing when and by whom information is accessed, helping youth feel secure in how their data is handled. Features such as customizable visibility settings, emotion-based check-ins, and reflective prompts can help youth express concerns with online risks in a supportive space. Ultimately, by facilitating open, non-judgmental communication between, these systems can provide youth a confidential and safe place to express themselves, ask questions, and empower them to seek help without fear of punishment, making trust a foundation of digital safety.

\subsubsection{Designing for Relational Healing and Online-to-Offline Support}

Online safety technologies for youth in the CWS should center therapeutic support and relational care through emotionally intelligent, trusted digital companions, as visioned by GALs. Research shows that such companions, including AI-embedded Non-Player Characters (NPCs), can provide empathy, validation, and gentle guidance over time \cite{baffa2017dealing, pretty2024case}, while offering therapeutic scaffolding that supports mental health and emotion regulation \cite{gao2024potential, silva2024gamemotion}. By simulating affective responses, these companions can help youth practice empathy, build self-awareness, and feel supported during emotional distress, elements which GALs noted as essential for youth in the CWS. Designed with trauma-informed principles in mind \cite{zhang2021trauma, kramer2013statewide}, such systems can foster emotional resilience, guide youth through online risks, and strengthen their capacity for meaningful connection both online and offline. Importantly, these technologies should also facilitate healthy transitions into offline routines and relationships, which GALs emphasized as vital for helping youth rebuild trust and form stable connections in their daily lives. Features that prompt screen breaks, encourage journaling, reading, or checking in with trusted adults can bridge digital engagement with offline healing and interpersonal development. By further combining therapeutic support with offline activation, these systems align more closely with youths’ developmental and relational needs, offering a holistic and trauma-informed alternative to surveillance-driven approaches.

However, designing AI companions for vulnerable populations like youth in the CWS requires a careful consideration of potential ethical issues \cite{kaimara2022could}. Current responsible and trustworthy AI guidelines are not tailored to marginalized populations \cite{abdulai2025generative}, therefore, it is especially important to design AI-based support systems with care to address the unique vulnerabilities of youth in the CWS. While these systems must adhere to ethical AI practices, such as transparent data collection, informed consent and assent, as well as algorithmic accountability to protect youths' privacy and prevent bias, it is crucial to build them around trauma-informed frameworks to ensure that these experiences do not re-traumatize youth \cite{shaik2025leveraging, scott2023trauma}. As pointed out by GALs, it is crucial to make sure AI-embedded experiences augment not substitute human relationships, avoiding youth developing full reliance on the AI system.

%\cite{mark2017invitation}
%Key Implications for Practice

% Trauma-informed assessment: Understanding online behaviors within the context of relational trauma rather than individual pathology
% Relationship-first interventions: Prioritizing trust-building and emotional safety before addressing specific online behaviors
% Collaborative safety planning: Involving youth as partners in developing their own online safety strategies
% Systemic support: Ensuring that multiple adults in youth's lives can provide consistent, caring guidance rather than fragmented oversight
% Long-term perspective: Recognizing that building the relational capacity for online safety is a developmental process that requires patience and sustained commitment

\subsection{Limitations and Future Work}

While this study offers insights for reconceptualizing online safety in the CWS through a systemic, trauma-informed lens and identifies key design principles for supporting vulnerable youth, it has several limitations. First, our workshop involved GALs from a single county in Indiana and included only one session, limiting the generalizability and depth of our findings. Future work should engage GALs across diverse jurisdictions and conduct multiple workshops to capture a wider range of experiences and explore their proposed solutions more deeply. Second, our participant sample was predominantly older GALs, whose perspectives and technological comfort levels may differ from those of younger or more digitally savvy professionals. Expanding participation to include GALs with varied demographics and technological backgrounds could yield richer insights. Third, this study focused exclusively on GALs as a stakeholder group. While their advocacy role provides critical information, future research should also engage directly with youth in the CWS, as well as therapists, biological parents, and other caregivers, to build a more comprehensive understanding of the online safety landscape and co-design interventions that reflect multiple perspectives. Future work should further prioritize youth-centered participatory approaches, ensuring that solutions are grounded in the lived experiences of youth themselves. Additionally, research should move toward implementing and evaluating prototypes informed by the systemic, trust-building principles identified here, testing their feasibility and effectiveness both in controlled and real-world settings. Finally, scholars and practitioners should explore how legal frameworks, court orders, and inter-agency coordination can be reformed to support collaborative, empowerment-oriented online safety policies that move beyond punitive restrictions toward holistic, systemic solutions.

%trauma-informed, empowerment-centered design principles
%NPCs (Non-Player Characters)
\section{Conclusion}

Our research highlights the urgent need to redesign online safety interventions for youth in the CWS by moving beyond restrictive models. Through co-design with GALs, we identified key shortcomings in current online safety of these youth, challenges GALs face with advocating for them, and collaboratively explored technology solutions that prioritize trust, therapeutic support, and multi-stakeholder coordination. By framing online safety as a matter of relational care, we emphasize that supporting youth wellbeing online cannot be disentangled from the broader ecosystems of support that shape their lives. This work invites HCI/GROUP communities to take a more holistic, equity-driven approach to designing for vulnerable youth, approach that centers care, collaboration, and youth voice in every layer of technological intervention.

%\begin{acks}

%\end{acks}

\balance
\bibliographystyle{ACM-Reference-Format}
\bibliography{00_refs}

@article{kensing1998participatory,
  title={Participatory design: Issues and concerns},
  author={Kensing, Finn and Blomberg, Jeanette},
  journal={Computer supported cooperative work (CSCW)},
  volume={7},
  pages={167--185},
  year={1998},
  publisher={Springer}
}

@book{schon2017reflective,
  title={The reflective practitioner: How professionals think in action},
  author={Sch{\"o}n, Donald A},
  year={2017},
  publisher={Routledge}
}

@article{gudowsky2017into,
  title={Into blue skies—a transdisciplinary foresight and co-creation method for adding robustness to visioneering},
  author={Gudowsky, Niklas and Sotoudeh, Mahshid},
  journal={NanoEthics},
  volume={11},
  pages={93--106},
  year={2017},
  publisher={Springer}
}

@misc{The_GAL_2019,
  author       = {National CASA/GAL Association for Children},
  title        = {The CASA/GAL Model},
  howpublished = {Web page, National CASA/GAL Association for Children},
  year         = {2019},
  url          = {https://nationalcasagal.org/our-work/the-casa-gal-model/},
  note         = {Last accessed 10-27-2025}
}

@misc{GAL_Role_2012,
  author       = {Fostering Perspectives},
  title        = {Guardians ad Litem Play a Vital Role},
  year         = {2012},
  url          = {https://fosteringperspectives.org/fpv17n1/GALs.htm},
  note         = {Last accessed 10-27-2025}
}

@misc{FAUHow2025Gal,
author = {Gisele Galoustian},
  title = {FAU | How Florida’s Guardian ad Litems Build Trust with Youth in Foster Care},
  url = "https://www.fau.edu/newsdesk/articles/foster-care-study",
month = {Apr},
year = {2025},
}

@incollection{Badillo-Urquiola2017abandoned,
	author = {Badillo-Urquiola, Karla and Harpin, Scott and Wisniewski, Pamela},
	title = {{Abandoned but Not Forgotten: Providing Access While Protecting Foster Youth from Online Risks}},
	booktitle = {{ACM Conferences}},
	pages = {17--26},
	year = {2017},
	month = jun,
	publisher = {Association for Computing Machinery},
	address = {New York, NY, USA},
	doi = {10.1145/3078072.3079724}
}

@inproceedings{oguine2026define_safety,
  author    = {Oguine, Ozioma Collins and Alvarado Garcia, Adriana and Muller, Michael and Badillo-Urquiola, Karla},
  title     = {{Who Gets to Define Safety? A Systematic Review of How Generative AI Research Addresses Youth Online Safety}},
  booktitle = {{ACM Conferences}},
  year      = {2026},
  month      = Apr,
  location  = {Barcelona, Spain},
  publisher = {Association for Computing Machinery},
  address   = {New York, NY, USA},
  doi       = {10.1145/3772318.3791346}
}

@incollection{Tanni2024May,
	author = {Tanni, Tangila Islam and Akter, Mamtaj and Anderson, Joshua and Amon, Mary Jean and Wisniewski, Pamela J.},
	title = {{Examining the Unique Online Risk Experiences and Mental Health Outcomes of LGBTQ+ versus Heterosexual Youth}},
	booktitle = {{ACM Conferences}},
	pages = {1--21},
	year = {2024},
	month = may,
	publisher = {Association for Computing Machinery},
	address = {New York, NY, USA},
	doi = {10.1145/3613904.3642509}
}

@incollection{Badillo2018stakeholder,
	author = {Badillo-Urquiola, Karla and Abraham, Jaclyn and Ghosh, Arup Kumar and Wisniewski, Pamela},
	title = {{A Stakeholders' Analysis of the Systems that Support Foster Care}},
	booktitle = {{ACM Conferences}},
	pages = {158--161},
	year = {2018},
	month = jan,
	publisher = {Association for Computing Machinery},
	address = {New York, NY, USA},
	doi = {10.1145/3148330.3154521}
}

@article{badillo2024caseworker,
author = {Badillo-Urquiola, Karla and Agha, Zainab and Abaquita, Denielle and Harpin, Scott B. and Wisniewski, Pamela J.},
title = {Towards a Social Ecological Approach to Supporting Caseworkers in Promoting the Online Safety of Youth in Foster Care},
year = {2024},
issue_date = {April 2024},
publisher = {Association for Computing Machinery},
address = {New York, NY, USA},
volume = {8},
number = {CSCW1},
url = {https://doi-org.proxy.library.nd.edu/10.1145/3637412},
doi = {10.1145/3637412},
journal = {Proc. ACM Hum.-Comput. Interact.},
month = apr,
articleno = {135},
numpages = {28}
}

@misc{AFcars,
author = {AFCARS},
  title = {Data and Statistics: AFCARS | The Administration for Children and Families},
  url = "https://acf.gov/cb/research-data-technology/statistics-research/afcars",
month = may,
year = {2025},
}

@inproceedings{badillo2019risk,
author = {Badillo-Urquiola, Karla and Page, Xinru and Wisniewski, Pamela J.},
title = {Risk vs. Restriction: The Tension between Providing a Sense of Normalcy and Keeping Foster Teens Safe Online},
year = {2019},
isbn = {9781450359702},
publisher = {Association for Computing Machinery},
address = {New York, NY, USA},
url = {https://doi-org.proxy.library.nd.edu/10.1145/3290605.3300497},
doi = {10.1145/3290605.3300497},
booktitle = {Proceedings of the 2019 CHI Conference on Human Factors in Computing Systems},
pages = {1–14},
numpages = {14},
location = {Glasgow, Scotland Uk},
series = {CHI '19}
}

@article{oguine2025Chins,
author = {Oguine, Ozioma C. and Park, Jinkyung Katie and Akter, Mamtaj and Olesk, Johanna and Alluhidan, Abdulmalik and Wisniewski, Pamela and Badillo-Urquiola, Karla},
title = {How the Internet Facilitates Adverse Childhood Experiences for Youth Who Self-Identify as in Need of Services},
year = {2025},
issue_date = {May 2025},
publisher = {Association for Computing Machinery},
address = {New York, NY, USA},
volume = {9},
number = {2},
url = {https://doi-org.proxy.library.nd.edu/10.1145/3710995},
doi = {10.1145/3710995},
journal = {Proc. ACM Hum.-Comput. Interact.},
month = may,
articleno = {CSCW097},
numpages = {39}
}

@inproceedings{saxena2023analysis,
author = {Saxena, Devansh and Moon, Erina Seh-Young and Chaurasia, Aryan and Guan, Yixin and Guha, Shion},
title = {Rethinking "Risk" in Algorithmic Systems Through A Computational Narrative Analysis of Casenotes in Child-Welfare},
year = {2023},
isbn = {9781450394215},
publisher = {Association for Computing Machinery},
address = {New York, NY, USA},
url = {https://doi-org.proxy.library.nd.edu/10.1145/3544548.3581308},
doi = {10.1145/3544548.3581308},
booktitle = {Proceedings of the 2023 CHI Conference on Human Factors in Computing Systems},
articleno = {873},
numpages = {19},
location = {Hamburg, Germany},
series = {CHI '23}
}

@inproceedings{saxena2023public_sector,
author = {Saxena, Devansh},
title = {Designing Human-Centered Algorithms for the Public Sector A Case Study of the U.S. Child-Welfare System},
year = {2023},
isbn = {9781450399456},
publisher = {Association for Computing Machinery},
address = {New York, NY, USA},
url = {https://doi-org.proxy.library.nd.edu/10.1145/3565967.3571759},
doi = {10.1145/3565967.3571759},
booktitle = {Companion Proceedings of the 2023 ACM International Conference on Supporting Group Work},
pages = {66–68},
numpages = {3},
location = {Hilton Head, SC, USA},
series = {GROUP '23}
}

@inproceedings{saxena2020group,
author = {Saxena, Devansh and Badillo-Urquiola, Karla and Wisniewski, Pamela and Guha, Shion},
title = {Child Welfare System: Interaction of Policy, Practice and Algorithms},
year = {2020},
isbn = {9781450367677},
publisher = {Association for Computing Machinery},
address = {New York, NY, USA},
url = {https://doi-org.proxy.library.nd.edu/10.1145/3323994.3369888},
doi = {10.1145/3323994.3369888},
booktitle = {Companion Proceedings of the 2020 ACM International Conference on Supporting Group Work},
pages = {119–122},
numpages = {4},
location = {Sanibel Island, Florida, USA},
series = {GROUP '20}
}

@inproceedings{oguine2024mainstream,
author = {Oguine, Ozioma Collins and Anuyah, Oghenemaro and Hughes, Emelia M. and Badillo-Urquiola, Karla},
title = {Examining Mainstream News Media Narratives on Youth Online Safety},
year = {2024},
isbn = {9798400711145},
publisher = {Association for Computing Machinery},
address = {New York, NY, USA},
url = {https://doi.org/10.1145/3678884.3681874},
doi = {10.1145/3678884.3681874},
booktitle = {Companion Publication of the 2024 Conference on Computer-Supported Cooperative Work and Social Computing},
pages = {349–354},
numpages = {6},
location = {San Jose, Costa Rica},
series = {CSCW Companion '24}
}

@article{saxena2024street_level,
author = {Saxena, Devansh and Guha, Shion},
title = {Algorithmic Harms in Child Welfare: Uncertainties in Practice, Organization, and Street-level Decision-making},
year = {2024},
issue_date = {March 2024},
publisher = {Association for Computing Machinery},
address = {New York, NY, USA},
volume = {1},
number = {1},
url = {https://doi-org.proxy.library.nd.edu/10.1145/3616473},
doi = {10.1145/3616473},
journal = {ACM J. Responsib. Comput.},
month = mar,
articleno = {2},
numpages = {32}
}

@inproceedings{saxena2020pd_study,
author = {Saxena, Devansh and Guha, Shion},
title = {Conducting Participatory Design to Improve Algorithms in Public Services: Lessons and Challenges},
year = {2020},
isbn = {9781450380591},
publisher = {Association for Computing Machinery},
address = {New York, NY, USA},
url = {https://doi-org.proxy.library.nd.edu/10.1145/3406865.3418331},
doi = {10.1145/3406865.3418331},
booktitle = {Companion Publication of the 2020 Conference on Computer Supported Cooperative Work and Social Computing},
pages = {383–388},
numpages = {6},
location = {Virtual Event, USA},
series = {CSCW '20 Companion}
}

@inproceedings{zytko2022participatory,
author = {Zytko, Douglas and J. Wisniewski, Pamela and Guha, Shion and P. S. Baumer, Eric and Lee, Min Kyung},
title = {Participatory Design of AI Systems: Opportunities and Challenges Across Diverse Users, Relationships, and Application Domains},
year = {2022},
isbn = {9781450391566},
publisher = {Association for Computing Machinery},
address = {New York, NY, USA},
url = {https://doi-org.proxy.library.nd.edu/10.1145/3491101.3516506},
doi = {10.1145/3491101.3516506},
booktitle = {Extended Abstracts of the 2022 CHI Conference on Human Factors in Computing Systems},
articleno = {154},
numpages = {4},
location = {New Orleans, LA, USA},
series = {CHI EA '22}
}

@article{caddle2022challenge,
author = {Caddle, Xavier V. and Naher, Nurun and Miller, Zachary P. and Badillo-Urquiola, Karla and Wisniewski, Pamela J.},
title = {Duty to Respond: The Challenges Social Service Providers Face When Charged with Keeping Youth Safe Online},
year = {2022},
issue_date = {January 2023},
publisher = {Association for Computing Machinery},
address = {New York, NY, USA},
volume = {7},
number = {GROUP},
url = {https://doi-org.proxy.library.nd.edu/10.1145/3567556},
doi = {10.1145/3567556},
journal = {Proc. ACM Hum.-Comput. Interact.},
month = dec,
articleno = {6},
numpages = {35}
}

@inproceedings{brown2019Algo,
author = {Brown, Anna and Chouldechova, Alexandra and Putnam-Hornstein, Emily and Tobin, Andrew and Vaithianathan, Rhema},
title = {Toward Algorithmic Accountability in Public Services: A Qualitative Study of Affected Community Perspectives on Algorithmic Decision-making in Child Welfare Services},
year = {2019},
isbn = {9781450359702},
publisher = {Association for Computing Machinery},
address = {New York, NY, USA},
url = {https://doi-org.proxy.library.nd.edu/10.1145/3290605.3300271},
doi = {10.1145/3290605.3300271},
booktitle = {Proceedings of the 2019 CHI Conference on Human Factors in Computing Systems},
pages = {1–12},
numpages = {12},
location = {Glasgow, Scotland Uk},
series = {CHI '19}
}

@inproceedings{McNally2018codesign,
author = {McNally, Brenna and Kumar, Priya and Hordatt, Chelsea and Mauriello, Matthew Louis and Naik, Shalmali and Norooz, Leyla and Shorter, Alazandra and Golub, Evan and Druin, Allison},
title = {Co-designing Mobile Online Safety Applications with Children},
year = {2018},
isbn = {9781450356206},
publisher = {Association for Computing Machinery},
address = {New York, NY, USA},
url = {https://doi-org.proxy.library.nd.edu/10.1145/3173574.3174097},
doi = {10.1145/3173574.3174097},
booktitle = {Proceedings of the 2018 CHI Conference on Human Factors in Computing Systems},
pages = {1–9},
numpages = {9},
location = {Montreal QC, Canada},
series = {CHI '18}
}

@inproceedings{oguine2025multistakeholder,
author = {Oguine, Ozioma C. and Olesk, Johanna and Solyst, Jaemarie and Madaio, Michael and Muller, Michael and Alvarado Garcia, Adriana and Badillo-Urquiola, Karla},
title = {Bridging Expertise and Participation in AI: Multistakeholder Approaches to Safer AI Systems for Youth Online Safety},
year = {2025},
isbn = {9798400714801},
publisher = {Association for Computing Machinery},
address = {New York, NY, USA},
url = {https://doi-org.proxy.library.nd.edu/10.1145/3715070.3748294},
doi = {10.1145/3715070.3748294},
booktitle = {Companion Publication of the 2025 Conference on Computer-Supported Cooperative Work and Social Computing},
pages = {150–155},
numpages = {6},
location = {
},
series = {CSCW Companion '25}
}

@inproceedings{cheng2022Workers,
author = {Cheng, Hao-Fei and Stapleton, Logan and Kawakami, Anna and Sivaraman, Venkatesh and Cheng, Yanghuidi and Qing, Diana and Perer, Adam and Holstein, Kenneth and Wu, Zhiwei Steven and Zhu, Haiyi},
title = {How Child Welfare Workers Reduce Racial Disparities in Algorithmic Decisions},
year = {2022},
isbn = {9781450391573},
publisher = {Association for Computing Machinery},
address = {New York, NY, USA},
url = {https://doi-org.proxy.library.nd.edu/10.1145/3491102.3501831},
doi = {10.1145/3491102.3501831},
booktitle = {Proceedings of the 2022 CHI Conference on Human Factors in Computing Systems},
articleno = {162},
numpages = {22},
location = {New Orleans, LA, USA},
series = {CHI '22}
}

@inproceedings{agha2022realtime,
author = {Agha, Zainab and Zhang, Zinan and Obajemu, Oluwatomisin and Shirley, Luke and J. Wisniewski, Pamela},
title = {A Case Study on User Experience Bootcamps with Teens to Co-Design Real-Time Online Safety Interventions},
year = {2022},
isbn = {9781450391566},
publisher = {Association for Computing Machinery},
address = {New York, NY, USA},
url = {https://doi-org.proxy.library.nd.edu/10.1145/3491101.3503563},
doi = {10.1145/3491101.3503563},
booktitle = {Extended Abstracts of the 2022 CHI Conference on Human Factors in Computing Systems},
articleno = {40},
numpages = {8},
location = {New Orleans, LA, USA},
series = {CHI EA '22}
}

@article{fitch2012youth,
  title={Youth in foster care and social media: A framework for developing privacy guidelines},
  author={Fitch, Dale},
  journal={Journal of Technology in Human Services},
  volume={30},
  number={2},
  pages={94--108},
  year={2012},
  publisher={Taylor \& Francis}
}

@article{gustavsson2015positive,
  title={Positive Youth Development and foster care youth: A digital perspective},
  author={Gustavsson, Nora and MacEachron, Ann},
  journal={Journal of Human Behavior in the Social Environment},
  volume={25},
  number={5},
  pages={407--415},
  year={2015},
  publisher={Taylor \& Francis}
}

@article{oguine2023you,
  title={You Don't Belong Here: Ableist Microaggressions on Adolescents with Disability (ies) and Special Needs in Social Virtual Reality (VR)},
  author={Oguine, Ozioma Collins and Badillo-Urquiola, Karla},
  journal={Available at SSRN 4381787},
  year={2023}
}

@inproceedings{multistakeholder2025,
author = {Oguine, Ozioma C. and Olesk, Johanna and Solyst, Jaemarie and Madaio, Michael and Muller, Michael and Alvarado Garcia, Adriana and Badillo-Urquiola, Karla},
title = {Bridging Expertise and Participation in AI: Multistakeholder Approaches to Safer AI Systems for Youth Online Safety},
year = {2025},
isbn = {9798400714801},
publisher = {Association for Computing Machinery},
address = {New York, NY, USA},
url = {https://doi-org.proxy.library.nd.edu/10.1145/3715070.3748294},
doi = {10.1145/3715070.3748294},
booktitle = {Companion Publication of the 2025 Conference on Computer-Supported Cooperative Work and Social Computing},
pages = {150–155},
numpages = {6},
location = {
},
series = {CSCW Companion '25}
}

@article{gomez2016shifting,
  title={Shifting the focus: Nonpathologizing approaches to healing from betrayal trauma through an emphasis on relational care},
  author={G{\'o}mez, Jennifer M and Lewis, Jenn K and Noll, Laura K and Smidt, Alec M and Birrell, Pamela J},
  journal={Journal of Trauma \& Dissociation},
  volume={17},
  number={2},
  pages={165--185},
  year={2016},
  publisher={Taylor \& Francis}
}

@article{purvis2013trust,
  title={Trust-based relational intervention (TBRI): A systemic approach to complex developmental trauma},
  author={Purvis, Karyn B and Cross, David R and Dansereau, Donald F and Parris, Sheri R},
  journal={Child \& youth services},
  volume={34},
  number={4},
  pages={360--386},
  year={2013},
  publisher={Taylor \& Francis}
}

@article{ball2021agency,
  title={Agency, genuine support, and emotional connection: Experiences that promote relational permanency in foster care},
  author={Ball, Barbara and Sevillano, Lalaine and Faulkner, Monica and Belseth, Tymothy},
  journal={Children and Youth Services Review},
  volume={121},
  pages={105852},
  year={2021},
  publisher={Elsevier}
}

@misc{munford2022children,
  title={Children and young people in the care system: Relational practice in working with transitions and challenges},
  author={Munford, Robyn},
  journal={Australian social work},
  volume={75},
  number={1},
  pages={1--4},
  year={2022},
  publisher={Taylor \& Francis}
}

@inproceedings{ghosh2020circle,
  title={Circle of trust: a new approach to mobile online safety for families},
  author={Ghosh, Arup Kumar and Hughes, Charles E and Wisniewski, Pamela J},
  booktitle={Proceedings of the 2020 CHI Conference on Human Factors in Computing Systems},
  pages={1--14},
  year={2020}
}

@article{rutkowski2021family,
  title={Family communication: examining the differing perceptions of parents and teens regarding online safety communication},
  author={Rutkowski, Tara L and Hartikainen, Heidi and Richards, Kirsten E and Wisniewski, Pamela J},
  journal={Proceedings of the ACM on Human-Computer Interaction},
  volume={5},
  number={CSCW2},
  pages={1--23},
  year={2021},
  publisher={ACM New York, NY, USA}
}

@article{caddle2025building,
  title={Building a Village: A Multi-stakeholder Approach to Open Innovation and Shared Governance to Promote Youth Online Safety},
  author={Caddle, Xavier V and Qadir, Sarvech and Hughes, Charles and Sweigart, Elizabeth A and Park, Jinkyung Katie and Wisniewski, Pamela J},
  journal={arXiv preprint arXiv:2504.03971},
  year={2025}
}

@article{akter2022parental,
  title={From parental control to joint family oversight: Can parents and teens manage mobile online safety and privacy as equals?},
  author={Akter, Mamtaj and Godfrey, Amy J and Kropczynski, Jess and Lipford, Heather R and Wisniewski, Pamela J},
  journal={Proceedings of the ACM on Human-Computer Interaction},
  volume={6},
  number={CSCW1},
  pages={1--28},
  year={2022},
  publisher={ACM New York, NY, USA}
}

@article{mark2017invitation,
  title={An invitation to internet safety and ethics: School and family collaboration.},
  author={Mark, Lauren K and Nguyen, Thanh Truc T},
  journal={Journal of Invitational Theory and Practice},
  volume={23},
  pages={62--75},
  year={2017},
  publisher={ERIC}
}

@inproceedings{qadir2024towards,
  title={Towards a Safer Digital Future: Exploring Stakeholder Perspectives on Creating a Sustainable Youth Online Safety Community},
  author={Qadir, Sarvech and Niser, Andy and Caddle, Xavier V and Alsoubai, Ashwaq and Park, Jinkyung Katie and Wisniewski, Pamela J},
  booktitle={Extended Abstracts of the CHI Conference on Human Factors in Computing Systems},
  pages={1--10},
  year={2024}
}

@article{glisson1998effects,
  title={The effects of organizational climate and interorganizational coordination on the quality and outcomes of children’s service systems},
  author={Glisson, Charles and Hemmelgarn, Anthony},
  journal={Child abuse \& neglect},
  volume={22},
  number={5},
  pages={401--421},
  year={1998},
  publisher={Elsevier}
}

@inproceedings{baffa2017dealing,
  title={Dealing with the emotions of non player characters},
  author={Baffa, Augusto and Sampaio, Pedro and Feij{\'o}, Bruno and Lana, Mauricio},
  booktitle={2017 16th Brazilian Symposium on Computer Games and Digital Entertainment (SBGames)},
  pages={76--87},
  year={2017},
  organization={IEEE}
}

@inproceedings{gao2024potential,
  title={The Potential and Mechanism of Artificial Intelligence Driven Non-Player Characters in Video Games for Anxiety Management},
  author={Gao, Yuzhen and Hu, Zhehan and Liu, Junhao},
  booktitle={2024 International Conference on Artificial Intelligence and Communication (ICAIC 2024)},
  pages={353--362},
  year={2024},
  organization={Atlantis Press}
}

@article{pretty2024case,
  title={A case for personalized non-player character companion design},
  author={Pretty, Emma J and Fayek, Haytham M and Zambetta, Fabio},
  journal={International Journal of Human--Computer Interaction},
  volume={40},
  number={12},
  pages={3051--3070},
  year={2024},
  publisher={Taylor \& Francis}
}

@inproceedings{silva2024gamemotion,
  title={GamEmotion: a serious game for emotion regulation in young adolescents},
  author={Silva, Eliana and Fran{\c{c}}a, Pedro and Reis, Lu{\'\i}s Paulo},
  booktitle={2024 IEEE conference on games (CoG)},
  pages={1--4},
  year={2024},
  organization={IEEE}
}

@article{darlington2005practice,
  title={Practice challenges at the intersection of child protection and mental health},
  author={Darlington, Yvonne and Feeney, Judith A and Rixon, Kylie},
  journal={Child \& Family Social Work},
  volume={10},
  number={3},
  pages={239--247},
  year={2005},
  publisher={Wiley Online Library}
}

@article{leathers2009context,
  title={Context-specific mental health services for children in foster care},
  author={Leathers, Sonya J and Atkins, Marc S and Spielfogel, Jill E and McMeel, Lorri S and Wesley, Julia M and Davis, Rafe},
  journal={Children and Youth Services Review},
  volume={31},
  number={12},
  pages={1289--1297},
  year={2009},
  publisher={Elsevier}
}

@article{hwang2017information,
  title={Information sharing between the child welfare and behavioral health systems: Perspectives from four stakeholder groups},
  author={Hwang, Sophia HJ and Mollen, Cynthia J and Kellom, Katherine S and Dougherty, Susan L and Noonan, Kathleen G},
  journal={Social Work in Mental Health},
  volume={15},
  number={5},
  pages={500--523},
  year={2017},
  publisher={Taylor \& Francis}
}

@article{lee2015accessing,
  title={Accessing quality early care and education for children in child welfare: Stakeholders' perspectives on barriers and opportunities for interagency collaboration},
  author={Lee, Sei-Young and Benson, Stephanie M and Klein, Sacha M and Franke, Todd M},
  journal={Children and Youth Services Review},
  volume={55},
  pages={170--181},
  year={2015},
  publisher={Elsevier}
}

@article{akin2017successes,
  title={Successes and challenges in developing trauma-informed child welfare systems: A real-world case study of exploration and initial implementation},
  author={Akin, Becci A and Strolin-Goltzman, Jessica and Collins-Camargo, Crystal},
  journal={Children and Youth Services Review},
  volume={82},
  pages={42--52},
  year={2017},
  publisher={Elsevier}
}

@article{kaimara2022could,
  title={Could virtual reality applications pose real risks to children and adolescents? A systematic review of ethical issues and concerns},
  author={Kaimara, Polyxeni and Oikonomou, Andreas and Deliyannis, Ioannis},
  journal={Virtual reality},
  volume={26},
  number={2},
  pages={697--735},
  year={2022},
  publisher={Springer}
}

@phdthesis{marie2025foster,
  title={Foster Children’s Repeated Disruptions: A Qualitative Analysis of Stakeholder Insights on Failed Permanency in Foster Care},
  author={Marie, Teresa},
  year={2025},
  school={University of Nevada, Las Vegas}
}

@article{edwards2023administrative,
  title={Administrative burdens in child welfare systems},
  author={Edwards, Frank and Fong, Kelley and Copeland, Victoria and Raz, Mical and Dettlaff, Alan},
  journal={RSF: The Russell Sage Foundation Journal of the Social Sciences},
  volume={9},
  number={5},
  pages={214--231},
  year={2023},
  publisher={RSF: The Russell Sage Foundation Journal of the Social Sciences}
}

@article{chuang2010role,
  title={The role of inter-agency collaboration in facilitating receipt of behavioral health services for youth involved with child welfare and juvenile justice},
  author={Chuang, Emmeline and Wells, Rebecca},
  journal={Children and youth services review},
  volume={32},
  number={12},
  pages={1814--1822},
  year={2010},
  publisher={Elsevier}
}

@article{smith2022collective,
  title={A collective impact approach to supporting youth transitioning out of government care},
  author={Smith, Annie and Peled, Maya and Martin, Stephanie},
  journal={Child abuse \& neglect},
  volume={130},
  pages={105104},
  year={2022},
  publisher={Elsevier}
}

@article{beyerlein2014need,
  title={Need for trauma-informed care within the foster care system},
  author={Beyerlein, Brittany A and Bloch, Ellin},
  journal={Child Welfare},
  volume={93},
  number={3},
  pages={7--22},
  year={2014},
  publisher={JSTOR}
}

@article{zhang2021trauma,
  title={Trauma-informed care for children involved with the child welfare system: A meta-analysis},
  author={Zhang, Saijun and Conner, Austin and Lim, Younghee and Lefmann, Tess},
  journal={Child Abuse \& Neglect},
  volume={122},
  pages={105296},
  year={2021},
  publisher={Elsevier}
}

@inproceedings{badillo2017understanding,
  title={Understanding the unique online challenges faced by teens in the foster care system},
  author={Badillo-Urquiola, Karla A and Ghosh, Arup Kumar and Wisniewski, Pamela},
  booktitle={Companion of the 2017 ACM Conference on Computer Supported Cooperative Work and Social Computing},
  pages={139--142},
  year={2017}
}

@article{sage2022systematic,
  title={A systematic review of internet communication technology use by youth in foster care},
  author={Sage, Melanie and Jackson, Sebrena},
  journal={Child and Adolescent Social Work Journal},
  volume={39},
  number={4},
  pages={375--390},
  year={2022},
  publisher={Springer}
}

@article{kumar2025cultivating,
  title={Cultivating a Supportive Sphere: Designing Technology to Increase Social Support for Foster-Involved Youth},
  author={Kumar, Ila K and Ferguson, Craig and Wu, Jiayi and Picard, Rosalind W},
  journal={Proceedings of the ACM on Human-Computer Interaction},
  volume={9},
  number={2},
  pages={1--24},
  year={2025},
  publisher={ACM New York, NY, USA}
}

@incollection{zhang2025dangerous,
  title={Dangerous Playgrounds: Child Players’ Encounters with Design-Mediated Risks on User Generated Game Platforms and Their Safety Practices},
  author={Zhang, Zinan and Gui, Xinning and Yu, Junnan and Bai, Sunhye and Kou, Yubo},
  booktitle={Proceedings of the 24th Interaction Design and Children},
  pages={296--313},
  year={2025}
}

@article{alford2019use,
  title={Use of smartphone technology in foster care to build relational competence: Voices of caregivers and implications for prudent parenting},
  author={Alford, Keith A and Denby, Ramona W and Gomez, Efren},
  journal={Journal of Family Social Work},
  volume={22},
  number={3},
  pages={209--230},
  year={2019},
  publisher={Taylor \& Francis}
}

@article{denby2016promoting,
  title={Promoting well-being through relationship building: The role of smartphone technology in foster care},
  author={Denby, Ramona W and Gomez, Efren and Alford, Keith A},
  journal={Journal of Technology in Human Services},
  volume={34},
  number={2},
  pages={183--208},
  year={2016},
  publisher={Taylor \& Francis}
}

@article{kramer2013statewide,
  title={A statewide introduction of trauma-informed care in a child welfare system},
  author={Kramer, Teresa L and Sigel, Benjamin A and Conners-Burrow, Nikki A and Savary, Patricia E and Tempel, Ashley},
  journal={Children and Youth Services Review},
  volume={35},
  number={1},
  pages={19--24},
  year={2013},
  publisher={Elsevier}
}

@inproceedings{park2024personally,
  title={Personally Targeted Risk vs. Humor: How Online Risk Perceptions of Youth vs. Third-Party Annotators Differ based on Privately Shared Media on Instagram},
  author={Park, Jinkyung and Gracie, Joshua and Alsoubai, Ashwaq and Razi, Afsaneh and Wisniewski, Pamela J},
  booktitle={Proceedings of the 23rd Annual ACM Interaction Design and Children Conference},
  pages={1--13},
  year={2024}
}

@article{akter2025calculating,
  title={Calculating Connection vs. Risk: Understanding How Youth Negotiate Digital Privacy and Security with Peers Online},
  author={Akter, Mamtaj and Park, Jinkyung Katie and Headrick, Campbell Robinson and Page, Xinru and Wisniewski, Pamela J},
  journal={Proceedings of the ACM on Human-Computer Interaction},
  volume={9},
  number={7},
  pages={1--26},
  year={2025},
  publisher={ACM New York, NY, USA}
}

@inproceedings{alsoubai2022mosafely,
  title={MOSafely, Is that Sus? A Youth-Centric Online Risk Assessment Dashboard},
  author={Alsoubai, Ashwaq and Caddle, Xavier V and Doherty, Ryan and Koehler, Alexandra Taylor and Sanchez, Estefania and De Choudhury, Munmun and Wisniewski, Pamela J},
  booktitle={Companion Publication of the 2022 Conference on Computer Supported Cooperative Work and Social Computing},
  pages={197--200},
  year={2022}
}

@article{razi2023sliding,
  title={Sliding into My DMs: Detecting Uncomfortable or Unsafe Sexual Risk Experiences within Instagram Direct Messages Grounded in the Perspective of Youth},
  author={Razi, Afsaneh and AlSoubai, Ashwaq and Kim, Seunghyun and Ali, Shiza and Stringhini, Gianluca and De Choudhury, Munmun and Wisniewski, Pamela J},
  journal={Proceedings of the ACM on Human-Computer Interaction},
  volume={7},
  number={CSCW1},
  pages={1--29},
  year={2023},
  publisher={ACM New York, NY, USA}
}

@article{agha2023strike,
  title={" Strike at the Root": Co-designing Real-Time Social Media Interventions for Adolescent Online Risk Prevention},
  author={Agha, Zainab and Badillo-Urquiola, Karla and Wisniewski, Pamela J},
  journal={Proceedings of the ACM on Human-Computer Interaction},
  volume={7},
  number={CSCW1},
  pages={1--32},
  year={2023},
  publisher={ACM New York, NY, USA}
}

@inproceedings{iftikhar2021designing,
  title={Designing Parental Monitoring and Control Technology: A Systematic Review},
  author={Iftikhar, Zainab and Haq, Qutaiba Rohan ul and Younus, Osama and Sardar, Taha and Arif, Hammad and Javed, Mobin and Shahid, Suleman},
  booktitle={Human-Computer Interaction--INTERACT 2021: 18th IFIP TC 13 International Conference, Bari, Italy, August 30--September 3, 2021, Proceedings, Part IV 18},
  pages={676--700},
  year={2021},
  organization={Springer}
}

@inproceedings{hartikainen2016should,
  title={Should we design for control, trust or involvement? A discourses survey about children's online safety},
  author={Hartikainen, Heidi and Iivari, Netta and Kinnula, Marianne},
  booktitle={Proceedings of the The 15th International Conference on Interaction Design and Children},
  pages={367--378},
  year={2016}
}

@inproceedings{wisniewski2017parental,
  title={Parental control vs. teen self-regulation: Is there a middle ground for mobile online safety?},
  author={Wisniewski, Pamela and Ghosh, Arup Kumar and Xu, Heng and Rosson, Mary Beth and Carroll, John M},
  booktitle={Proceedings of the 2017 ACM Conference on Computer Supported Cooperative Work and Social Computing},
  pages={51--69},
  year={2017}
}

@ARTICLE{katieResiliance2024,
  author={Park, Jinkyung Katie and Akter, Mamtaj and Wisniewski, Pamela and Badillo-Urquiola, Karla},
  journal={IEEE Security \& Privacy}, 
  title={It’s Still Complicated: From Privacy-Invasive Parental Control to Teen-Centric Solutions for Digital Resilience}, 
  year={2024},
  volume={},
  number={},
  pages={2-12},
  doi={10.1109/MSEC.2024.3417804}
}

@article{caddle2023duty,
  title={Duty to Respond: The Challenges Social Service Providers Face When Charged with Keeping Youth Safe Online},
  author={Caddle, Xavier V and Naher, Nurun and Miller, Zachary P and Badillo-Urquiola, Karla and Wisniewski, Pamela J},
  journal={Proceedings of the ACM on Human-Computer Interaction},
  volume={7},
  number={GROUP},
  pages={1--35},
  year={2023},
  publisher={ACM New York, NY, USA}
}

@article{livingstone2008taking,
  title={Taking risky opportunities in youthful content creation: teenagers' use of social networking sites for intimacy, privacy and self-expression},
  author={Livingstone, Sonia},
  journal={New media \& society},
  volume={10},
  number={3},
  pages={393--411},
  year={2008},
  publisher={Sage Publications Sage UK: London, England}
}

@inproceedings{wisniewski2016dear,
  title={Dear diary: Teens reflect on their weekly online risk experiences},
  author={Wisniewski, Pamela and Xu, Heng and Rosson, Mary Beth and Perkins, Daniel F and Carroll, John M},
  booktitle={Proceedings of the 2016 CHI Conference on Human Factors in Computing Systems},
  pages={3919--3930},
  year={2016}
}

@inproceedings{agha2021just,
  title={‘Just-in-time’parenting: A two-month examination of the bi-directional influences between parental mediation and adolescent online risk exposure},
  author={Agha, Zainab and Ghaiumy Anaraky, Reza and Badillo-Urquiola, Karla and McHugh, Bridget and Wisniewski, Pamela},
  booktitle={International Conference on Human-computer interaction},
  pages={261--280},
  year={2021},
  organization={Springer}
}

@inproceedings{schiano2017parental,
  title={Parental controls: Oxymoron and design opportunity},
  author={Schiano, Diane J and Burg, Christine},
  booktitle={HCI International 2017--Posters' Extended Abstracts: 19th International Conference, HCI International 2017, Vancouver, BC, Canada, July 9--14, 2017, Proceedings, Part II 19},
  pages={645--652},
  year={2017},
  organization={Springer}
}

@inproceedings{badillo2019stranger,
  title={Stranger danger! social media app features co-designed with children to keep them safe online},
  author={Badillo-Urquiola, Karla and Smriti, Diva and McNally, Brenna and Golub, Evan and Bonsignore, Elizabeth and Wisniewski, Pamela J},
  booktitle={Proceedings of the 18th ACM international conference on interaction design and children},
  pages={394--406},
  year={2019}
}

@misc{cortesi2025frontiers,
  author       = {Sandra Cortesi and Urs Gasser},
  title        = {Frontiers in Digital Child Safety: Designing Child-Centered Digital Ecosystems that Support Rights, Agency, and Well-Being},
  institution  = {TUM Think Tank / Berkman Klein Center},
  year         = {2025},
  url          = {https://tumthinktank.de/wp-content/uploads/FRONTIERS-IN-DIGITAL-CHILD-SAFETY.pdf},
  note         = {Last accessed 11-02-2025}
}

@article{braun2021thematic,
  title={Thematic analysis: A practical guide},
  author={Braun, Virginia and Clarke, Victoria},
  year={2021},
  publisher={SAGE publications Ltd}
}

@article{abdulai2025generative,
  title={Is Generative AI Increasing the Risk for Technology-Mediated Trauma Among Vulnerable Populations?},
  author={Abdulai, Abdul-Fatawu},
  journal={Nursing Inquiry},
  volume={32},
  number={1},
  pages={e12686},
  year={2025},
  publisher={Wiley Online Library}
}

@article{shaik2025leveraging,
  title={Leveraging Data Science for Resilience: Improving Trauma-Informed Care Practice for Adverse Childhood Experience with AI \& Data Science Application},
  author={Shaik, Mohmmad Arif},
  year={2025}
}

@inproceedings{scott2023trauma,
  title={Trauma-informed social media: Towards solutions for reducing and healing online harm},
  author={Scott, Carol F and Marcu, Gabriela and Anderson, Riana Elyse and Newman, Mark W and Schoenebeck, Sarita},
  booktitle={Proceedings of the 2023 CHI Conference on Human Factors in Computing Systems},
  pages={1--20},
  year={2023}
}

@inproceedings{pinter2017adolescent,
  title={Adolescent online safety: Moving beyond formative evaluations to designing solutions for the future},
  author={Pinter, Anthony T and Wisniewski, Pamela J and Xu, Heng and Rosson, Mary Beth and Caroll, Jack M},
  booktitle={Proceedings of the 2017 Conference on Interaction Design and Children},
  pages={352--357},
  year={2017}
}

@article{oguine2025online,
  title={Online safety for all: Sociocultural insights from a systematic review of youth online safety in the global south},
  author={Oguine, Ozioma Collins and Anuyah, Oghenemaro and Agha, Zainab and Melgarez, Iris and Alvarado Garcia, Adriana and Badillo-Urquiola, Karla},
  journal={Proceedings of the ACM on Human-Computer Interaction},
  volume={9},
  number={7},
  pages={1--30},
  year={2025},
  publisher={ACM New York, NY, USA}
}

@article{freed2025protect,
  title={PROTECT: A Framework to Foster Digital Resilience for Youth Navigating Technology-Facilitated Abuse},
  author={Freed, Diana and Bazarova, Natalie and Consolvo, Sunny and Cosley, Dan and Gage Kelley, Patrick},
  journal={Social Sciences},
  volume={14},
  number={6},
  pages={378},
  year={2025},
  publisher={MDPI}
}

@inproceedings{agha2024tricky,
  title={Tricky vs. transparent: Towards an ecologically valid and safe approach for evaluating online safety nudges for teens},
  author={Agha, Zainab and Park, Jinkyung and Wan, Ruyuan and Ali, Naima Samreen and Wang, Yiwei and Difranzo, Dominic and Badillo-Urquiola, Karla and Wisniewski, Pamela J},
  booktitle={Proceedings of the 2024 CHI Conference on Human Factors in Computing Systems},
  pages={1--20},
  year={2024}
}

@article{thompson2025forming,
  title={Forming Relationships with Youth in Foster Care: Perspectives of Guardian ad Litems},
  author={Thompson, Heather M and Cooley, Morgan E and Cesar, Gabriel T and Backstrom, Laura and Colvin, Marianna L},
  journal={Journal of Child and Family Studies},
  volume={34},
  number={1},
  pages={68--82},
  year={2025},
  publisher={Springer}
}

%\appendix

\end{document}